\documentclass[twocolumn,nofootinbib,floatfix]{revtex4}
\usepackage{multirow}
\usepackage{fancyhdr}
\usepackage{array}
\usepackage[normalem]{ulem} 
\usepackage{amsfonts}
\usepackage{amsmath}
\usepackage{subfigure}
\usepackage{afterpage}
\usepackage{xspace}
\usepackage{color}
\usepackage{cancel}
\usepackage{xcolor}
\usepackage{graphicx}
\usepackage{textcomp}

\usepackage[letterpaper]{geometry}
\definecolor{linkcolor}{cmyk}{0,0.7,0.5,0.5}
\usepackage[colorlinks=true, pdfstartview=FitV, linkcolor=linkcolor, citecolor=blue, urlcolor=blue, bookmarksdepth=subparagraph]{hyperref}

\let\oldsubsubsection = \subsubsection

\newcommand{\outlineblank}[1]{\phantomsection\addcontentsline{toc}{#1}{ }}
\begin{document}
\title{Quantitative cw Overhauser DNP Analysis of Hydration Dynamics}
\newcommand{\affilucsb}{\affiliation{Department of Chemistry and Biochemistry,
    University of California,
    Santa Barbara, CA}}
\author{John M. Franck}\affilucsb
\author{Anna Pavlova}\affilucsb
\author{Songi Han}\affilucsb
\newcommand{\ximyhigh}{0.33\xspace}
\newcommand{\xireferenceval}{0.33\xspace}
\newcommand{\tauformyhigh}{33.3\xspace}
\newcommand{\capillarydim}{0.6\mm i.d. 0.84\mm o.d. quartz}
\newcommand{\xifcrArmstrong}{0.36\xspace}
\newcommand{\xifcrBennati}{0.33-0.35\xspace}
\newcommand{\xiTurke}{0.33\xspace} 
\newcommand{\ximd}{0.30\xspace}
\let\oldIm=\Im
\let\oldRe=\Re
\newcommand{\xiArmstrong}{0.22\xspace} 
\newcommand{\tauArmstrong}{76\ps\xspace} 
\newcommand{\taubulkequation}{\ensuremath{\taubulk=33.3\ps}\xspace} 
\newcommand{\Dlocal}{\ensuremath{D_{local}}\xspace}%
\newcommand{\correltime}{\ensuremath{\tau_c}\xspace}%
\newcommand{\myreftenmicromolar}{a\xspace}
\newcommand{\myrefonefiftymicromolar}{b\xspace}
\newcommand{\myreftwelvefiftymicromolar}{c\xspace}
\newcommand{\myrefhundredmillimolar}{d\xspace}
\newcommand{\ie}{i.e.\xspace}
\newcommand{\oht}{4-hydroxy-TEMPO\xspace}
\bibliographystyle{unsrt}
\newcommand{\etal}{et.~al.\xspace}
\newcommand{\Emax}{\ensuremath{E_{max}}\xspace}
\def\uM{\ensuremath{\;\mbox{\textmu M}}\xspace}
\def\uL{\ensuremath{\;\mbox{\textmu L}}\xspace}
\def\mM{\ensuremath{\;\mbox{mM}}\xspace}
\def\mL{\ensuremath{\;\mbox{mL}}\xspace}
\def\cm{\ensuremath{\;\mbox{cm}}\xspace}
\newcommand{\Real}[1]{\Re\mathfrak{e}\left[ #1 \right]}
\newcommand{\Imag}[1]{\Im\mathfrak{m}\left[ #1 \right]}
\newcommand\magn[1]{\ensuremath{\times 10^{#1}}\xspace}
\newcommand\Ang{\ensuremath{\;\mbox{\AA}}\xspace}
\newcommand\GHz{\ensuremath{\;\mbox{GHz}}\xspace}
\newcommand\MHz{\ensuremath{\;\mbox{MHz}}\xspace}
\newcommand\kHz{\ensuremath{\;\mbox{kHz}}\xspace}
\newcommand\mm{\ensuremath{\;\mbox{mm}}\xspace}
\newcommand\W{\ensuremath{\;\mbox{W}}\xspace}
\newcommand\K{\ensuremath{\;\mbox{K}}\xspace}
\def\M{\ensuremath{\;\mbox{M}}\xspace}
\def\nm{\ensuremath{\;\mbox{nm}}\xspace}
\def\m{\ensuremath{\;\mbox{m}}\xspace}
\newcommand\secs{\ensuremath{\;\mbox{s}}\xspace}
\newcommand\us{\ensuremath{\;\mbox{s}}\xspace}
\newcommand\ns{\ensuremath{\;\mbox{ns}}\xspace}
\newcommand\ps{\ensuremath{\;\mbox{ps}}\xspace}
\def\smax{\texorpdfstring{\ensuremath{s_{max}}}{smax}}
\def\tonen{\texorpdfstring{\ensuremath{T_{1,0}}}{T10}}
\newcommand\textsubscript[1]{\ensuremath{{}_{\mbox{#1}}}}
\newcommand{\obpair}[2]{\ensuremath{\overbrace{#1}^{\mbox{\begin{small}#2\end{small}}}}}
\pagestyle{fancy}
\makeatletter
  \fancyhf{}
  \lhead{J.M. Franck, A. Pavlova, and S. Han}
  \rhead{{\tiny \@title}\quad\thepage}
\makeatother
\begin{abstract}
Liquid state Overhauser Effect Dynamic Nuclear Polarization (ODNP)
has experienced a recent resurgence of interest.
In particular,
    a new manifestation
    of the ODNP measurement~\cite{Armstrong_jacs}
    measures
    the translational mobility of water within 5-10\Ang
    of an ESR-active spin probe
    (\ie the local translational diffusivity$D_{local}$ near an electron spin resonance
    active molecule).
Such spin probes, typically stable nitroxide radicals,
    have been
    attached to the surface or interior
    of macromolecules,
    including proteins~\cite{Pavlova_pccp,Armstrong_apomb},
    polymers~\cite{Ortony_njp},
    and membrane vesicles~\cite{Kausik_pccp}.
Despite the unique specificity of this measurement,
    it requires only
    a standard
    X-band
    ($\sim$10\GHz) continuous wave (cw)
    electron spin resonance (ESR) spectrometer,
    coupled with a standard nuclear magnetic resonance (NMR) spectrometer.
Here, we present a set of developments and corrections
    that allow us to improve the accuracy of quantitative ODNP and apply
    it to samples more than two orders of magnitude lower
    than were previously feasible.

An existing model
    for ODNP signal enhancements~\cite{Armstrong_jcp,Bates1977a,Hausser1968,HydeFreed}
    accurately predicts the ODNP enhancements for
    water that contains
    high ($\ge 10\mM $) concentrations of
    spin probes,
    whether they be freely dissolved in solution~\cite{Turke2010,Armstrong_jacs,Armstrong_jcp}
    or covalently tethered to slowly tumbling macromolecular systems~\cite{Armstrong_jacs,Ortony_njp}.
This model yields a parameter called the coupling factor, $\xi$,
which gives the efficiency of the ODNP polarization transfer
    in the presence of the spin label,
    and which depends only on the relative motion of the
    water molecules and the spin label.
Measurements of the ODNP enhancements and relaxation times can extract the parameter $\xi$, allowing one to read out the local translational dynamics
    of the water near the spin probe.
However,
    recent literature yields
    conflicting results
    for basic ODNP measurements
    of small spin probes dissolved in water~\cite{Armstrong_jacs,Armstrong_jcp,Turke2010,Turke_sat}
    and a closer inspection
    --
    especially at low concentrations of spin probes
    --
    reveals unexpected results that imply the breakdown of  the existing model as a result of microwave-induced sample heating.
Specifically, while the conventional model predicts that the enhancements should
    converge asymptotically to a maximum value, $E_{max}$,
    at high microwave powers,
    the enhancements instead continue to increase linearly.
In part due to this breakdown of the model,
    the concentration regime below $\sim$100\uM was
    previously quite infeasible for quantitative
    Overhauser DNP studies.

The technique
    presented here feasibly quantifies the ODNP coupling factor
    at lower concentrations
    by separately determining the two
    fundamental relaxivities involved in ODNP:
    the local cross-relaxivity, $k_\sigma$,
    and the local self-relaxivity, $k_\rho$,
    whose ratio gives the coupling factor, $\xi = k_\sigma/k_\rho$.
These relaxivities determine the
    concentration-dependent relaxation rates
    for the cross relaxation from the electrons to the protons,
    and for the self-relaxation from the protons
    near the spin probe to the bath (\ie ``lattice''),
    respectively. 
Enhancement vs. power ($E(p)$) curves
    acquired on cw ODNP instrumentation
    can quantify
    the cross-relaxivity ($k_\sigma$) for concentrations as low as tens of
    micromolar.
Furthermore, such data can include 
    a correction for the microwave heating effects
    previously mentioned.
Independent measurements can provide accurate values for the self-relaxivity
    ($k_\rho$) that are not affected by microwave heating,
    and which will have even further improved accuracy when obtained
    from samples of larger volume or higher concentration.
The more accurate value for the coupling factor,
    $\xi$, that results from this new technique
    more reliably quantifies the local translational diffusivity, \Dlocal,
    near the spin probe
    and opens up the novel possibility of analyzing lower sample
    concentrations of $\le 100\uM$ that are
    critical for biomolecular studies.

To demonstrate these improvements and
    compare to recent results,
    we repeat careful measurements
    of the coupling factor ($\xi$)
    between a small nitroxide probe
    (\oht)
    and otherwise unperturbed bulk water,
    at both high and low
    spin probe concentrations.
At high concentrations,
    we measure a significantly higher extrapolated enhancement, \Emax,
    than was previously measured or
    predicted by solely cw ODNP-based work~\cite{Armstrong_jcp}.
At all concentrations,
    for the first time,
    the data measured by the cw ODNP instrumentation shown here
    agrees with the coupling factor
    values of
    \xifcrArmstrong~\cite{Armstrong_jacs},
    \xifcrBennati~\cite{Bennati_fcr},
    or \xiTurke~\cite{Turke_sat,Turke2010}
    that others have reported based on
    ODNP measurements augmented by
    FCR experiments
    and pulsed ESR experiments,
    or the value of \ximd predicted
    by molecular dynamics simulations~\cite{Sezer_xi}.
On the one hand,
    this observation resolves the debate revolving around
    the absolute value of the coupling factor between
    water and freely dissolved spin probes,
    which is an important reference
    value for the study of hydration water in biological
    and other macromolecular systems.
    Our data conclusively supports a values of \xireferenceval~\cite{Turke_sat,Turke2010}
    rather than \xiArmstrong~\cite{Armstrong_jacs,Armstrong_jcp}.
On the other hand,
    contrary to conclusions drawn
    in previous literature~\cite{Turke_sat,HoferBennati_jacs},
    this data implies that
    solely cw ODNP methods can provide
    quantitative and accurate
    coupling factors, and thus
    derive accurate hydration dynamics information.
This is fortuitous;
    FCR and pulsed ESR tools will continue to present
    powerful and complementary capabilities,
    while the implementation of quantitative
    ODNP measurements on widely available
    and easy to use cw ODNP instrumentation
    has distinctly practical benefits for the end user.
\end{abstract}
\maketitle

\quad

\quad

\section{Introduction }
\outlineblank{subsection}
\outlineblank{subsubsection}

Overhauser-effect dynamic nuclear polarization (ODNP)
    can achieve the hyperpolarization of nuclear spins
    in aqueous solutions at ambient temperatures.

It requires only the addition of
    molecules or moieties containing unpaired electron spins
    (\ie~spin probes)
    and the significant saturation of
    their electron spin resonance transitions
    with resonant microwave irradiation~\cite{Lingwood_annrev}
    in order
    to increase the NMR signal of a sample solution
    by up to two orders of magnitude~\cite{Kentgens_pccp,Turke2010}
    relative to thermal polarization.
However,
    its capabilities far exceed
    just the efficient signal amplification of room temperature solutions.
Through even meager amplification of proton NMR signal of water,
    ODNP also provides an unprecedented measurement of local hydration dynamics,
    specifically quantifying the local diffusivity of water
    within only 5-10\Ang (2-4 layers of water)
    around a spin probe.
Since well established chemistry can attach
    stable nitroxide radical-based spin probes
    at arbitrarily chosen sites
    on proteins,
    lipid vesicles,
    synthetic polymers,
    and
    nucleic acids~\cite{Qin_firstattachment,Qin_natproto},
    ODNP can target the local hydration dynamics
    near a variety of sites,
    which can reside either within the core or
    on the surface of proteins or macromolecular assemblies~\cite{Armstrong_apomb,Armstrong_jacs,Pavlova_pccp,Ortony_njp}.

Two variants of ODNP have been reported:
    one which retrieves the necessary information from the NMR signal while relying solely on
    the use of a cw microwave source
    that saturates the ESR transition (as in \cite{Armstrong_jcp,Armstrong_jacs}),
    and one which relies at least partially on the ability to apply microwave
    pulses and detect the resulting ESR free induction decay or spin echo
    (as in \cite{Turke2010}).
The pairing of ODNP with
    pulsed ESR has shown promise by
    rectifying the value for the coupling
    factor between water and the free spin probe,
    and
    yielding results that agree with the predictions
    of FCR~\cite{Armstrong_jacs,Turke_sat}
    and MD~\cite{Sezer_xi} studies.
However, cw ODNP (\ie relying only on cw ESR instrumentation)
    demonstrates complementary capabilities. 
Various recent studies have shown the promise
    of the less expensive
    and more accessible cw ODNP variant for the
    determination of hydration dynamics~\cite{Armstrong_jacs,Armstrong_apomb,Pavlova_pccp}.
This is fortunate since many researchers
    and research facilities only have access
    to cw ESR instrumentation.
In fact, thus far,
    only the cw ODNP method has been applied to
    the quantification of hydration water dynamics in biological
    and soft matter systems.
However,
    a controversy over the accuracy of cw ODNP has persisted
    due to the fact that it
    was believed to report
    a value of the coupling factor
    of water near freely dissolved nitroxides~\cite{Armstrong_jacs}
    that disagreed
    with the value predicted from
    FCR measurements or the value observed by ODNP
    in combination with pulsed ESR.
This contrast validates investigations
    into the improvement
    of the accuracy and reproducibility of the cw ODNP method.

The understanding of the physical processes underlying ODNP enhancement
    has remained relatively unchanged since
    Hausser and Stehlik~\cite{Hausser1968}
    explained how the steady-state solution
    of the Solomon equations~\cite{Solomon1955}
    could predict ODNP enhancements.
Since then,
    researchers have applied this theory in a relatively unmodified form~\cite{Turke2010,Armstrong_jacs}
    by extending the models to predict
    the influence of electron spin saturation~\cite{Armstrong_jcp,Bates1977a},
    or integrating earlier models~\cite{HydeFreed,YinHyde,PoppHyde}
    that assist in directly measuring the electron spin saturation~\cite{Turke_sat}.
More recent models
    of high field ODNP
    have included the effects of sample heating
    in order to predict the enhancements of free spin probes
    with the purpose of achieving maximal signal
    enhancements~\cite{VanBentum_Kentgens}.

The predominant impact of microwave sample heating
    on the specific practical problem of extracting hydration dynamics, however,
    has not yet been elucidated or quantified.
At X-band frequencies (near 10\GHz and 3\cm wavelengths),
    the generation of a significant magnetic field (\ie~$B_1$)
    inside a finite-sized sample
    necessarily implies the generation of an electric field
    that will heat aqueous samples, even if to a small extent.
For instance, Bennati \etal~\cite{Turke2010} have directly observed such heating
    in very large samples ($\ge 0.9\mm$ ID)
    with an optical temperature sensor.
Such a measurement records temperature increases of up to 70$^o$C.
Their optical temperature sensor can not measure
    smaller diameter samples,
    which should exhibit less heating and therefore
    make more ideal ODNP samples.
Therefore, they estimate heating
    in the smaller diameter samples
    based on the observed
    dielectric losses,
    which they calculate from changes in the microwave cavity $Q$ factor.
For instance, 
    they use the change in cavity $Q$ factor
    to predict an increase of sample temperature
    of at least 20$^o$C for a
    sample with 0.45~mm diameter and 10~mm length.
Bennati \etal further demonstrated a procedure
    for minimizing temperature variation by carefully constraining
    the sample volume to the region of minimal electric field.
In their specific setup, they
    show negligible dielectric loss for a
    sample with 0.45~mm diameter and 3~mm length.
However, not all cw ESR setups can measure changes in $Q$ at high power,
    especially while providing the precision required
    to measure these dielectric losses.
This strategy may also overestimate the amount of sample heating,
    since it does not account for any heat transferred away from the
    sample and into the air that cools the cavity,
    which will become increasingly important with increased
    flow rates and decreased sample diameters.
Furthermore,
    the predictions based on dielectric losses
    and the measurements of the temperature sensor
    do not agree for microwave irradiation times longer than
    4\secs.
Moving forward, we should note that -- as a key requirement --
    hydration dynamics experiments
    call for an easily repeatable and verifiable
    measurement of sample heating
    that can be implemented with existing cw ESR and ODNP systems.

The study of biological systems typically
    requires lower (hundreds of\uM)
    concentrations of samples
    and consequently lower concentrations
    of the spin probes. 
One general effect we will present here is that
    even small sample heating can significantly
    lengthen the longitudinal relaxation time\footnote{We uniformly denote
        the bulk longitudinal relaxation time, {\it excluding}
        any spin probe-induced relaxation by $T_{1,0}$ and
        reserve $T_{1}$ to denote the relaxation times of samples
        that contain spin probes.} ($T_{1,0}$)
    of the bulk water -- \ie water in regions where
    dipolar interaction with the spin probe becomes insignificant.
The lengthening of $T_{1,0}$ impacts the reproducibility of
    ODNP measurements
    in several ways that
    were not previously anticipated,
    and is particularly significant at lower
    spin probe concentrations.

We demonstrate how
    the temperature variation of the bulk water $T_{1,0}$,
    which researchers characterized
    and modeled over 35 years ago~\cite{Hindman1973},
    does provide the most practically useful intrinsic
    probe of sample temperature
    in an ODNP experiment.
For instance, we demonstrate how this approach can easily track
    the relative quality of different ODNP probe
    designs, and we propose its application towards
    further advances in quantitative ODNP,
    through optimization of ODNP hardware and
    iterative temperature compensation.

Finally, we combine modest, but meaningful, hardware improvements
    with a new experimental procedure
    and data analysis method.
These advances both
    account for the lengthening of $T_{1,0}$ with increasing microwave power
    and help extract the
    cross-relaxivity, $k_\sigma$.
The separate extraction of $k_\sigma$
    specifically allows 
    the measurements of translational hydration dynamics
    at very low
    concentrations, as low as 10\uM,
    while all these advances
    yield visible
    improvement in the overall accuracy
    of the measurement of local hydration dynamics
    at both low and moderate concentrations of spin probe.

\section{Theory}
We begin by reviewing why the bulk water spin lattice ($T_1$) relaxation
    varies approximately linearly with temperature
    and discussing the physical origin
    of this change in temperature.
Then, after reviewing the current model for ODNP enhancements,
    we model how we can
    account for this change
    in bulk water relaxation with increasing microwave power.
This allows us to extract accurate and reproducible
    enhancement data and hydration dynamics results.
\subsection{$T_{1,0}$ is a Sensitive Probe of Sample Temperature}
\label{sec:intrinsic_probe}
We begin by reviewing
    a model for the temperature dependence
    of the NMR relaxation of pure water.
Note that, for consistency,
    we refer to the time constant for this
    relaxation as the $T_{1,0}$ time.
This is because the pure water treated in this section
    does not contain spin label.
The following sections will cover
    the relevance of this model
    and the resulting $T_{1,0}$ times to the
    $T_1$ times of sample solutions
    that contain spin probes.

Hindman \etal~\cite{Hindman1973}
    established a model that
    fits the experimentally observed $\tonen$ times
    across the full range of temperatures relevant to liquid water
    at atmospheric pressure.\footnote{For a more modern overview of
        MR thermometry, also see~\cite{Quesson2000}.}
It includes relaxations induced by fluctuations
    in the proton-proton dipolar interactions
    and by fluctuations in the spin rotational interactions
    (see also \cite{Green1965}).
Both the intermolecular and the intra-molecular dipolar
    contributions to the relaxation rate
    follow a temperature dependence consisting of the sum of two exponential terms,
    while the relaxation due to spin-rotational
    coupling varies directly with both temperature
    and the spin-rotational correlation time, $\tau_{sr}$.
Explicitly,
\begin{align}
    \frac{1}{\tonen} =&
    \obpair{A_1 e^{\frac{B_1}{T}} + A_2 e^{\frac{B_2}{T}}}{\parbox[c]{1in}{\centering nuclear-nuclear dipolar}}
    +\obpair{
    \frac{2}{9} k_B T h^{-2}
    \mbox{Tr}\left[ \mathbf{I} \right]
    \mbox{Tr}\left[ \mathbf{C}^2 \right]
    \tau_{sr}}{spin rotational},
    \label{eq:wood_rotational}
\end{align}
where $\mathbf{I}$ indicates the moment of inertia
    of the water molecules, which is 
    $\mbox{Tr}\left[ \mathbf{I} \right]=5.8783 \times 10^{-40}$~$\mbox{g}\cdot \cm ^2$,
    and $\mathbf{C}$
    indicates the spin-rotation interaction tensor,
    where 
    $\mbox{Tr}\left[ \mathbf{C}^2 \right] = 4\pi^2 1046.7\kHz^2$.
The weights of the two exponential terms that make up the dipolar
    relaxation are
    $A_1=4.6\times10^{-9}\secs^{-1}$
    and
    $A_2=6.3\times10^{-4}\secs^{-1}$,
    with associated exponential constants
    $B_1=4787\K$
    and
    $B_2=1764\K$  \cite{Hindman1973}.
Hindman's choice of 12.3\ps\ for the spin rotational
    coupling time, $\tau_{sr}$, fits the experimental data well.
Note that,
    as discussed by Hindman
    \etal~\cite{Hindman1973},
    $\tau_{sr}$
    is related to, but not numerically
    identical to,
    the rotational correlation times given
    by $^{17}O$ relaxation experiments.

\begin{figure}[tbp]
    \begin{center}
    \includegraphics[width=3.5in]{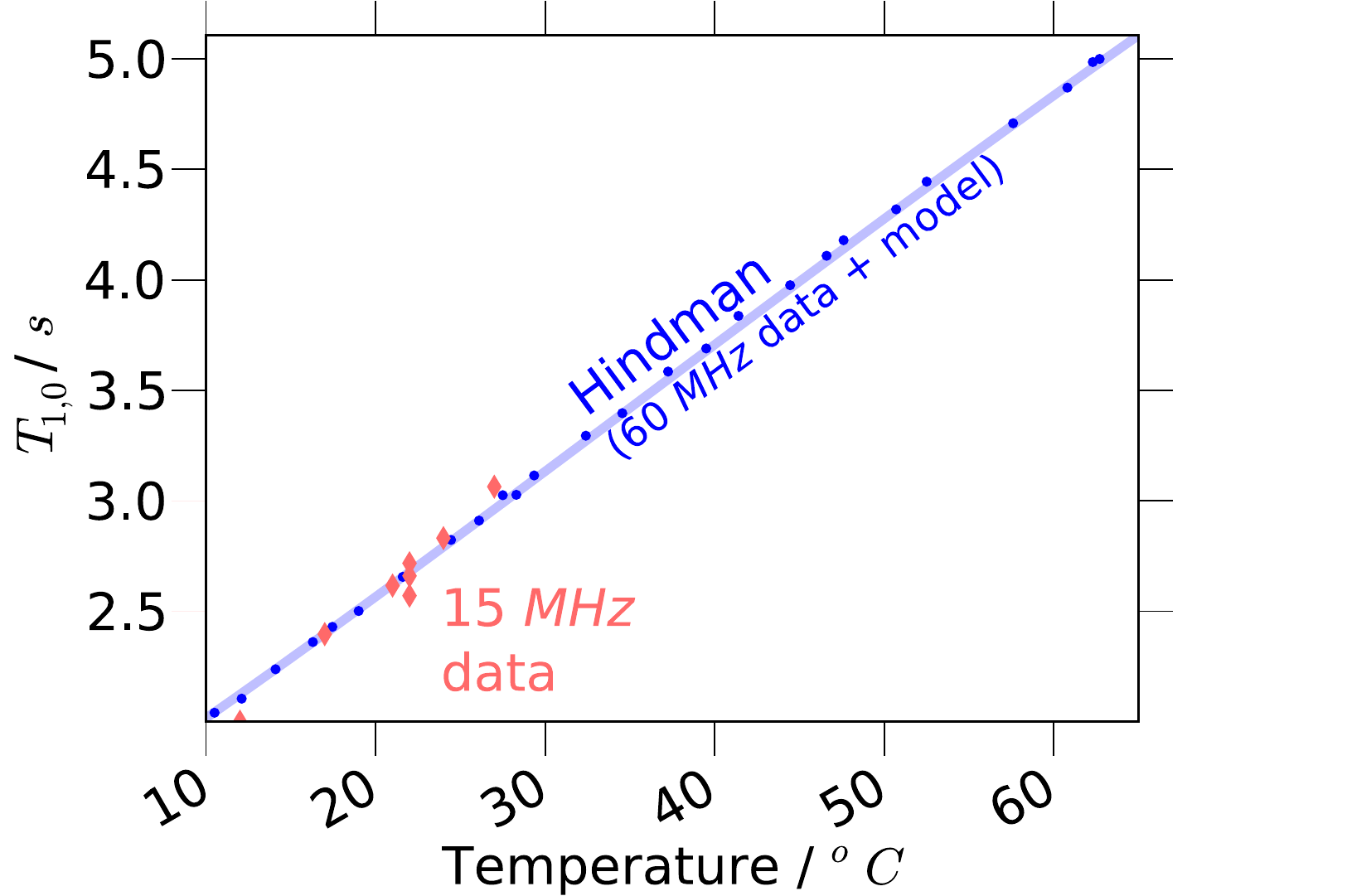}
    \end{center}
    \caption{Hindman's data (from \cite{Hindman1973}),
    for the longitudinal relaxation time, $\tonen$, of water
    vs. temperature, $T$, at an NMR resonance frequency of 60 \MHz,
    and the corresponding fit in blue (dark gray),
    accompanied by $T_{1,0}$ data
    that we measured at an NMR resonance frequency of 15~MHz
    at the sample conditions used throughout this paper in red (light gray).
    Since most applications of ODNP for the study of
    hydration dynamics will employ samples that are not deoxygenated,
        Hindman's data has been adjusted by increasing the relaxation rate by $70\times10^{-3}\secs^{-1}$ to account for the resulting additional relaxation mechanisms;
        the adjustment leads to a good match for the data we measured.
    (We denote the longitudinal relaxation time as $\tonen$ to explicitly exclude the presence of spin probe).}
    \label{fig:hindmantemp}
\end{figure}

Hindman's model points out that at tens of \MHz,
    proton Larmor frequencies fall in a regime where
    neither the dipolar nor the spin-rotational
    relaxation mechanism depends significantly
    on the magnetic field.
We measured several $T_{1,0}$ times with
    an ODNP probe in an ESR cryostat;
    these data fit well to the Hindman model,
    and imply the existence of a small additional
    relaxation contribution of $70\times10^{-3}\secs^{-1}$,
    which likely arises from the presence of standard
    amounts of oxygen in the 
    aqueous sample, unlike in Hindman's degassed water samples
    (fig.~\ref{fig:hindmantemp}).

The spin rotational component
    in eq.~\ref{eq:wood_rotational}
    only contributes significantly at temperatures approaching 100$^o$C.
Neglecting this component, the relaxation time,
    $T_{1,0}(T)$
    has a derivative maximum
    at 30$^o$C.
Importantly,
    (relative to the
    values of $B_1$ and $B_2$ of
    eq.~\ref{eq:wood_rotational})
    the range of temperatures
    relevant to ODNP studies of hydration dynamics
    (which generally probe hydration dynamics
    within $\pm 20^o$C of ambient
    temperature)
    falls near this derivative maximum.
Thus, the $T_{1,0}$ time
    responds dramatically to small changes in
    temperature,
 and induces correspondingly significant
    changes in the resulting ODNP enhancements.

This derivative maximum near ambient temperature
    makes the $T_{1,0}$ time
    of water
    an important intrinsic
    measurement of the sample temperature,
    which can track the changes in sample temperature
    with increasing microwave power.
Obviously, the $T_{1,0}$ time
    will be most sensitive to temperature
    changes near ambient temperature.
Furthermore, since the curvature (\ie second derivative w.r.t. temperature)
    vanishes, the $T_{1,0}$ time also
    depends approximately linearly on temperature
   (fig.~\ref{fig:hindmantemp}).

For standard measurements of the $T_1$ and $T_{1,0}$ times,
    we can fit the integrated signal intensities ($c M(\tau)$)
    from inversion recovery or saturation recovery
    experiments to the standard form
    \newcommand{\minfty}{\ensuremath{M(\infty)}}
    \begin{align}
        c M(\tau) &=
        c \minfty
        +
        c\left( M(0)-\minfty \right)
        e^{-\tau/T_1},
        \label{eq:t1fit}
    \end{align}
where $\tau$ gives the magnetization recovery delay,
    $M(0)$ gives the initial magnetization in the recovery curve
    ($-M(\infty)$ for inversion recovery,
    0 for saturation recovery),
    $M(\infty)$ gives the steady-state magnetization (\ie~after infinite recovery delay),
    and $c$ is a flexible fit parameter.
Note that the ``$T_1$'' above could equally well be
    the $T_1$ of a sample with spin probe,
    or the $T_{1,0}$ of one without.

We will also require a method
    capable of measuring
 time-dependent variations in the $T_{1,0}$ time
    that occur on a timescale faster than  $T_{1,0}$.
By rapidly repeating saturation-recovery with
    a fixed recovery time, $\tau$,
    much shorter than $T_{1,0}$,
    we can repeat a train of acquisitions
    at a rate of two to three scans per $T_{1,0}$ period,
    and so avoid the recovery time of
    $5\times T_{1,0}$ between subsequent signal acquisitions
    required by inversion recovery measurements.\footnote{Note the distinction here, between the repetition delay ($5\times T_1$) needed to recover magnetization between subsequent scans, and the recovery delay, $\tau$, which is either a fixed value (as here) or the indirect dimension of the $T_1$ experiment (as in the full inversion or saturation recovery experiment).}
Eq.~\ref{eq:t1fit} will still allow a relative comparison
    of $T_{1,0}$ times for the various acquisitions,
    even when $\tau$ is significantly
    small relative to $T_{1,0}$.
Though intrinsically less accurate than an
    inversion recovery
    or saturation recovery experiment acquired
    over several recovery delay points,
    the faster experiment will allow us to determine
    --
    with about 1~s time resolution
    --
    how rapidly the sample temperature responds
    to changes in incident microwave power.
\subsection{Source of the Microwave-Induced Heating Effect}

The interaction between the electric field of the microwaves
    and the aqueous solution (which is a dielectric)
    induces changes in bulk water dynamics,
    which lead to the changes in relaxation
    time, $T_{1,0}$,
    that we measure.
For convenience, our model will describe
    this change in dynamics
    simply as an increase in the temperature of the
    solution,
    \ie~as heating.
However,
    before we consider 
    the impact on the ODNP enhancements,
    we first examine
    the effects of the
    dielectric interaction.
In particular,
    we note that -- while sufficient within the scope of this article --
    a temperature-based description
    may provide only a limited
    insight into a more complex interaction.

Simply put,
    the electric field only induces specific
    modes of molecular motion.
Excitation of a mode can contribute a mixture of
    adiabatic and irreversible changes
    to the molecular dynamics within the sample solution.
Meanwhile,
    the air flowing around the capillary
    that contains the sample
    is actively removing heat from the system,
 even as heat is entering the system
    via the dielectric excitation.
To the best of our knowledge,
    it has not been clarified whether
    or not one expects
    all modes of molecular
    motion to maintain a thermal equilibrium 
    or not under such a setup.

We believe it is worthwhile to pause and
    clarify the effect of dielectric interactions
    in terms of the standard Debye model.
Though, in this article, we will not quantitatively employ this equation,
    it helps us to classify the timescales
    and molecular motions associated with
    the dielectric interaction between the
    microwaves and the sample solution.
Specifically, it gives a complex permittivity (\ie complex dielectric coefficient), $\hat{\varepsilon}(\omega)$,
    that varies with the frequency, $\omega$, of the electric field
    of the radiation, namely,
\begin{align}
    \hat{\varepsilon}(\omega)
    =&
    \varepsilon_{\infty}
    +
    \sum_k
    \frac{c_k}{1+i \omega \tau_{k}}
    , 
    \label{eq:debye}
\end{align}
    which is broken down
    in terms of $\varepsilon_\infty$
    -- the limiting dielectric permittivity
    at frequencies with periods far faster than the relevant
    relaxation times --
    and a sum over $k$
    dielectric relaxation processes (\ie mechanisms).
The $\tau_k$ are the dielectric relaxation times associated with
    the various relaxation mechanisms,
    the $c_k$ are the coefficients describing the relative
    dominance of the different mechanisms,
    and
    $\omega$ is the angular frequency ($\left[ \mbox{rad}/\mbox{s} \right]$) of the electric field
    of the radiation.

It is important to note that all dielectric interactions relevant
    for water at X-band (\ie~$\sim$10\GHz)
    frequencies necessarily involve changes to the molecular dynamics
    of the water.
In pure water,
    there is only one mechanism that contributes significantly
    to the dielectric permittivity
    in the range of frequencies up to and including X-band
    microwaves (\ie~$\sim$10\GHz).
The relevant molecular motions
    involve overall rotations
    of the water molecule about axes perpendicular
    to its electric dipole
    and require a finite
    relaxation time of $\tau_k = 8.3\ps$~\cite{Kaatze_water}.
Solutes introduced into the water can have various effects
    on this dielectric relaxation process:
    they may change the
    relaxation time, $\tau_k$, of the bulk water;
    they may inhibit the modes of motion that contribute to the 8.3\ps\
    process, leading to a decrease in the corresponding $c_k$ value;
    and/or -- especially in the case of macromolecules --
    they may introduce new relaxation processes
    with slower relaxation times,
    thus introducing new terms to the sum over
    the $k$ dielectric relaxation mechanisms aside
    from the one corresponding to the 8.3\ps\ process~\cite{Kaatze_water,Halle2004,oleinikova2004can}. 

By remembering that the microwave resonator (e.g.  cavity)
    is simply an electric circuit,
    we come to better understand
    the meaning and relevance
    of the complex permittivity.
A sample
    whose dielectric permittivity, $\hat{\varepsilon}(\omega)$,
    at the incident microwave frequency
    is entirely real-valued
    can sit in the electric field
    generated by the resonator 
    without dissipating any power
    and therefore without
    generating any heat.
Of course, the sample
    does change the
    capacitance (or, more generally, reactance)
    of the circuit because it cyclically stores
    and releases the energy of the applied
    electric field
    by changing the alignment of its
    molecular electric dipoles.
The real part
    of the dielectric permittivity thus quantifies how much
    ``reactive'' or ``adiabatic''\footnote{Note that this motion
        is ``adiabatic'' both in the sense that it is isothermal,
        and also slow relative to the capability of the water
        matrix to respond to the rotation.}
    molecular motion the electric field induces
    in the sample solution.
By contrast, a sample with an imaginary-valued dielectric
    implies an increase in the effective resistance of
    the resonator,
    which in turn implies a conversion of microwave power
    into heat.
Stated differently, the relative magnitude of the imaginary part
    gives the amount of incident radiation
    absorbed by the solution that is eventually dissipated
    as heat~\cite{JacksonChpt7}.
As this motion leads to a change in the
    heat in the system,
    unlike the molecular motion induced
    by the real part of the dielectric,
    we refer to it as ``irreversible.''

The form of eq.~\ref{eq:debye}
    allows us to identify
    whether
    the contribution of
    a particular dielectric relaxation mechanism
    (\ie~$k$ term)
    is significant at a particular microwave
    frequency, and if so,
    whether the associated molecular motions
    cause adiabatic or irreversible changes
    to the molecular dynamics.
Specifically, when the incident microwave frequency
    is much slower than
    the dielectric relaxation time ($\omega \ll 1/\tau_k$),
    $\hat{\varepsilon}$ is entirely real,
    corresponding to an adiabatic change in dynamics.
At angular frequencies near $1/\tau_k$
    the imaginary part
    of the $k^{th}$ mechanism
    (\ie $\left|\Imag{c_k /1 + i \omega \tau_k }\right| = c_k \omega \tau_k / 1 + \omega^2 \tau_k^2$) 
    rises to a broad maximum,
    which indicates irreversible changes in the dynamics.
For example, for the primary relaxation process in bulk water
    at 8.3\ps~\cite{Kaatze_water},
    the water molecules will respond adiabatically to fields
    in the DC to low microwave regime.
Over a broad range
    of frequencies near $1/2 \pi \tau_k \approx 19\GHz$,
    the sample absorbs microwaves,
    whose energy irreversibly drives
    the relatively disordered modes of motion
    associated with this relaxation process.
Simultaneously,
    at these frequencies,
    the real component of the dielectric
    (\ie $\Real{ c_k/1+i \omega \tau_k } = c_k / 1 + \omega^2 \tau_k^2$)
    and adiabatic changes to the dynamics
    coming from that particular mode (\ie the $k^{th}$ mode)
    fall to half their maximum amplitude,
    reaching zero at higher frequencies.
Finally, at even higher frequencies
    ($\omega \gg 1/\tau_k$),
    the imaginary component and associated irreversibly
    driven motion associated with that particular mode
    become negligible as well.

Thus, the electric fields in the X-band regime
    will change the molecular dynamics of the solution
    in several ways
    that could potentially alter the NMR
    bulk relaxation time, $T_{1,0}$.
In X-band ODNP experiments,
    we employ a microwave frequency of $\sim$10\GHz,
    which we expect will induce both adiabatic
    and irreversibly driven dipole reorientation
    in the bulk water.
Each could potentially contribute to changes of the $T_{1,0}$ time.
However, only the irreversibly driven dipole
    reorientation will change
    the actual sample temperature.
Hydration water on bimolecular surfaces
    display slower modes of motion
    that may further complicate this picture.
For instance, in a recent study of
    an aqueous solution of ribonuclease A
    at ambient temperature,
    Oleinikova~\etal
    present a dielectric relaxation
    time $\tau_k$ of 35\ps,
    which they assign to reorientation
    of the hydration water~\cite{oleinikova2004can}. 
We expect
    the dielectric interaction to
    irreversibly drive some limited changes
    to the molecular dynamics of such hydration water.
The extent to which these 
    changes impact the translational hydration water dynamics,
    as measured by ODNP,
    as well as the rate of molecular exchange
    and heat transfer between
    the hydration water and the bulk water
    could both provide interesting information,
    but remain relatively undetermined.

This understanding provides an important
    background for the way we use
    the word ``temperature'' in this article.
If we were to compare 
    one sample that is
    irradiated by microwaves at $\sim$10\GHz,
    while simultaneously actively cooled by air flowing
    around the sample capillary tube (like in our experimental setup),
    to another sample of identical composition
    that is not irradiated
    but, rather, simply heated until it yields the same $T_{1,0}$
    as the first sample,
then we must concede
    that the molecular dynamics
    of the two samples may not match.
Similarly,
    it is not trivial to determine what, if any,
    effect the adiabatically driven motion will have on the $T_{1,0}$ rate.
All these complications could lead to a richer
    interpretation of the heating effect that may validate future studies.
In particular,
    as previously explained (eq.~\ref{eq:wood_rotational}),
    the value of $T_{1,0}$ depends primarily on
    proton-proton dipolar coupling,
    which lends one to believe that it would be
    more sensitive to the rotations of water molecules
    induced by the dielectric interaction.
In fact,
    the temperature dependence of $T_{1,0}$
    has even been shown to closely follow that
    of the rotational motion of the individual water molecules,
    as -- for instance -- observed by $^{17}$O NMR~\cite{Hindman1973}.
This fact,
    in combination with the previously mentioned
    complications lends interesting motivations
    to direct future publications towards investigating
    the observation (presented later)
    that the relaxivities $k_\rho$ and $k_\sigma$,
    which encode the information about the local dynamics of water
    near the spin probe, remain relatively constant with
    microwave power, despite relatively dramatic changes
    in $T_{1,0}$.

Setting these interesting possibilities aside,
    however,
    throughout this article
    we employ the definition of effective ``temperature''
    to be the equilibrium temperature at which the $T_{1,0}$
    is elevated to the same value
    that we observe under steady-state dielectric
    excitation.
This is the definition most relevant to
    the changes in the
    ODNP measurements
    that we observe.
\subsection{Impact of Heating on ODNP}
\renewcommand{\subsubsection}[1]{\oldsubsubsection{#1}}
Through changes in the $T_{1,0}$ time,
    increases in the effective temperature drive
    increases in the steady-state ODNP enhancement,
    as well as increases in the repetition delay necessary between each
    NMR acquisition.
After reviewing the extant theory,
    which is derived predominantly from Hausser and Stehlik~\cite{Hausser1968},
    we first explain
    how lengthened NMR relaxation times
    at high power can lead to artifacts 
    in the ODNP enhancement vs. power, \ie $E(p)$,
    data.
Then we outline how the lengthened $T_{1,0}$ time
    leads to unexpectedly high values
    for the actual enhancement.
Finally, we present a new strategy for data analysis
    that both compensates for these changes in $T_{1,0}$
    and allows determination of the hydration dynamics
    at tens of micromolar concentrations.
\subsubsection{Review of Previous Theory}

For clarity, we note that while previous work by Han \etal~\cite{Armstrong_jacs,Armstrong_jcp,Pavlova_pccp,Ortony_njp,Kausik_jacs,Lingwood_annrev}
    has denoted the coupling factor with $\rho$,
    here we will use $\xi$ for the coupling factor,
    since (consistently with other literature~\cite{Turke2010,Hausser1968,Cavanagh_book,Abragam_book})
    we reserve $\rho$ for the local self-relaxation rate (table~\ref{tab:symbol_table_han}).

The polarization transferred from the electron spin to the protons
    is inverted relative
    to the thermal polarization
    of the protons.
Thus, sufficient
    amounts of polarization transfer lead
    to enhanced and inverted NMR signal.
The enhancement, $E$,
    is defined as
    the ratio between the ODNP-enhanced proton signal and the thermal
    signal.
We also find it convenient to refer to the
    amount of ``polarization transferred,''\footnote{$1-E$ gives the polarization
    transferred in units of thermal
    polarization of the proton.} $1-E$.
As shown by Hausser and Stehlik~\cite{Hausser1968},
    the amount of polarization transferred
    can be broken down into a product,
    \newcommand{\xsp}{\ensuremath{\xi s(p)}\xspace}%
    \newcommand{\ksp}{\ensuremath{k_\sigma s(p)}\xspace}%
    \newcommand{\ksm}{\ensuremath{k_\sigma s_{max}}\xspace}%
    \begin{align}
        1-E(p) = \xsp f \left| \frac{\omega_e}{\omega_H} \right|
        \label{eq:polarization}
        ,
    \end{align}
    where the coupling factor, $\xi$,
    the leakage factor, $f$,
    and the saturation factor, $s$,
    all affect the efficiency with which
    electron spin polarization,
    whose equilibrium population
    is approximately proportional to the
    ESR resonance frequency $\omega_e$,
    transfers to the inverted nuclear spin polarization,
    whose equilibrium population is
    approximately proportional
    to the proton Larmor
    frequency $\omega_H$.
    \newcommand\polratio{659.32\xspace}%
For nitroxide spin probes in aqueous solution,
    we have measured the ratio of $\omega_e$ to $\omega_H$
    as\footnote{In practical application,
        one can more accurately
        measure the relevant
        (microwave and radio) frequencies
        than one can measure the absolute value of the static field
        with a Hall probe.} \polratio.
To extract the information pertaining to local hydration dynamics,
    we wish to accurately isolate
    the coupling factor, $\xi$,
    from the other parameters.

The leakage factor, $f$, gives the proportion
    of the total proton relaxation
    that is due to the local dipolar\footnote{(\ie~induced
        by dipolar interaction with the electron spin)}
        relaxation mechanisms
\begin{equation}
    f =
    \frac{\rho}{\rho + T_{1,0}^{-1}}
    \label{eq:feq}
\end{equation}
Since the total
    longitudinal relaxation rate
    of a solution containing a spin probe, $T_1^{-1}$,
    is the sum of the bulk, $T_{1,0}^{-1}$,
    and the local dipolar relaxation rate, $\rho$,
    \ie
\begin{equation}
    T_1^{-1} = \rho + T_{1,0}^{-1}
    ,
    \label{eq:t1sum}
\end{equation}
one typically writes the leakage
    factor as
\begin{equation}
    f=  1-\frac{T_1}{T_{1,0}.}
    ,
    \label{eq:feq_exp}
\end{equation}
since inversion recovery experiments can directly determine
    both $T_1$ and $T_{1,0}$.

The saturation factor, $s(p)$,
    gives the net saturation of all electron spins in the sample.
In the absence of microwave power,
    $s=0$,
    while in the limit of microwave power,
    it approaches a value of $s_{max}$.
For $^{14}$N nitroxides, $\frac{1}{3}\le s_{max}\le 1$,
    as discussed below.
Standard ESR theory~\cite{Poole}, as well as its specific application
    to ODNP~\cite{Armstrong_jcp,Bates1977a}
    indicates that (for well separated Lorentzian absorption lines),
    the saturation factor follows an asymptotic form
    \newcommand\phalf{\ensuremath{p_{1/2}}\xspace}
    \begin{align}
        s(p)
        =&
        \frac{s_{max} p}{\phalf+p}
        .
        \label{eq:smax}
    \end{align}
Here, \phalf\ is the power needed to achieve half of the maximum
    possible saturation.
For clarity, we can substitute this into eq.~\ref{eq:polarization}, yielding
    \begin{align}
        1-E(p) = \xi f \left| \frac{\omega_e}{\omega_H} \right|
        \frac{s_{max} p}{\phalf+p}
        \label{eq:polarization_subst}
        .
    \end{align}
A variety of factors,
    including the electron spin relaxation time and the electrical
    properties of the resonant cavity
    combine to determine the value of \phalf.
Therefore, the previously employed analysis extrapolates
    the enhancements to their asymptotic limit
    \begin{align}
        \lim_{p\rightarrow \infty}E
        \equiv&E_{max}
        ,
        \intertext{thus eliminating the dependence on \phalf\ to give
        (via \ref{eq:polarization} and \ref{eq:smax})}
        1-E_{max}
        =&\xi s_{max} f \left|\frac{\omega_e}{\omega_H}\right|
        \label{eq:emax}
        .
    \end{align}
Later, for reasons that will become obvious,
    we will refer to
    eq.~\ref{eq:polarization_subst}-\ref{eq:emax}
    (where $f$ is assumed a constant)
    as the ``uncorrected model''
    when it is employed to experimentally determine the coupling factor.

To determine $\xi$ from \ref{eq:emax},
    one must still determine the value of $s_{max}$.
Previously, Armstrong and Han~\cite{Armstrong_jcp} presented
    the most complete analysis describing
    the key mechanisms behind total electron spin saturation
    (\ie the net contribution from all ESR lines).
Their analysis is relevant for quantifying the ODNP effect
    for both freely dissolved spin probes
    and those tethered to larger molecular systems,
    as it describes the effects of both Heisenberg exchange
    and nitrogen spin relaxation on
    the three (or two) hyperfine states of
    $^{14}$N (or $^{15}$N) of stable nitroxide spin probes
    in order to predict
    $s_{max}$ as a function of free nitroxide spin probe concentration
    and nitrogen nuclear spin relaxation.
Notably, they
    predict that for biological or polymer samples with
    covalently tethered spin probes, $s_{max}$
    should closely approach 1~\cite{Armstrong_jcp}.
    \begin{table}
        \newcommand\leftstyle[1]{\begin{flushright}#1\end{flushright}}
        \newcommand\csty[1]{\makebox[2cm][c]{#1}}
            \centering
            {\scriptsize
            \begin{tabular*}{\linewidth}{@{\extracolsep{\fill}}m{2cm}m{2cm}lm{2cm}}
\hline\hline
                                 & \csty{{\it Previous} Notation of Han \etal} &               & \csty{Standard Notation}\\
\hline
\leftstyle{self-relaxation rate} & \csty{{\color{gray} $k C$}}                 & $\Rightarrow$ & \csty{$\rho$}\\
\leftstyle{coupling factor}      & \csty{{\color{gray} $\rho$}}                & $\Rightarrow$ & \csty{$\xi$}\\
                \hline\hline
            \end{tabular*}
            }
            \caption{For those more familiar with notation
                previously used by Han et al., this table
                translates to the standard notation.}
            \label{tab:symbol_table_han}

            {\scriptsize
            \renewcommand{\arraystretch}{0}
            \newcommand\changerowspace{0pt}
            \begin{tabular*}{\linewidth}{@{\extracolsep{\fill}}m{2cm}ccc}
            \hline\hline
            \noalign{\smallskip}
                                          & {\scriptsize Standard Symbol} & {\scriptsize Relaxivities}                              & {\scriptsize Transition Rates} \\
 \noalign{\smallskip}\hline
 \leftstyle{local dipolar self-relaxation rate}   & $\rho$                        & $k_\rho C$                                              & {\scriptsize $w_0 + 2 w_1 + w_2$} \\ [\changerowspace]
 \leftstyle{local dipolar cross-relaxation rate}  & $\sigma$                      & $k_\sigma C$                                            & {\scriptsize $w_2 - w_0$} \\ [\changerowspace]
 \leftstyle{coupling factor}                      & $\xi$                         & $\displaystyle\frac{k_\sigma}{k_\rho}$                  & {\scriptsize $\displaystyle\frac{w_2 - w_0}{w_0 + 2 w_1 +w_2}$} \\ [\changerowspace]
 \leftstyle{leakage factor}                       & $f$                           & $\displaystyle\frac{k_\rho C}{k_\rho C + T_{1,0}^{-1}}$ & {\scriptsize $\displaystyle\frac{w_0 + 2 w_1 + w_2}{w_0 + 2 w_1 + w_2 + w^0}$}\\ [\changerowspace]
 \leftstyle{nuclear longitudinal relaxation rate} & $T_1^{-1}$                    & $k_\rho C + T_{1,0}^{-1}$                               & {\scriptsize $w_0 + 2 w_1 + w_2 + w^0$}\\
            \hline\hline
        \end{tabular*}
        \label{tab:symbol_table_compare}
        }
        \caption{Each column summarizes an equivalent means for denoting a quantity:
        the terminology for the parameter,
        the symbol used in the standard notation,
        and equivalent expressions in terms of
        both the concentration-independent relaxivities
        and
        the transition rates, respectively.
    Wherever possible, all notation matches that of
        Solomon~\cite{Solomon1955} and Hausser~\cite{Hausser1968}.
    This work, which explores various concentration-dependent effects,
        makes frequent use of the concentration-independent
        relaxivities
        (even though the $k_\rho$, $k_\sigma$ notation
        is not used elsewhere).
    Furthermore, since $k_\rho$
        and $k_\sigma$ explicitly denote relaxation processes
    localized (\ie~$\propto r^{-6}$) around the spin probe,
        they provide interesting information about the
        hydration dynamics,
        which is complementary to information
        derived from the coupling factor, $\xi$
        (compare, for instance to~\cite{Armstrong_apomb,HodgesBryant}).
    This publication does not make use of the transition
            rate notation~\cite{Hausser1968,Solomon1955},
            but it is provided here in the third column
            for comparison.}
        \label{tab:symbol_table}
    \end{table}

In summary,
    the previous (\ie uncorrected) analysis advises
    one to determine the enhancement at several
    values of microwave power
    and to extrapolate to an asymptotic limit of
    the signal enhancements,
    $E_{max}$, as a function of microwave power.
The value of $E_{max}$ is then
    inserted into eq.~\ref{eq:emax}.
By employing inversion recovery experiments with and without spin probe to
    obtain (respectively) $T_1$ and $T_{1,0}$ as inputs for eq.~\ref{eq:feq_exp},
    one could obtain a value for the leakage factor ($f$)
    needed for eq.~\ref{eq:emax}.
Finally, one can determine
    the value for $s_{max}$ in eq.~\ref{eq:emax}
    through one of several means:
    from a concentration series~\cite{Armstrong_jcp,Bates1977a},
    from calculations
    based on the known exchanges rates
    for small freely dissolved spin probes
    determined by Bennati \etal~\cite{Turke_sat},
    or from the reasonable approximation that
    $s_{max}\approx1$ for most macromolecular
    (e.g. proteins, polymers, vesicles)
    samples with tethered spin probes.
Researchers have employed the uncorrected analysis (eq.~\ref{eq:emax})
    fairly routinely to extract $\xi$.
By way of the translational correlation time, \correltime,
    given by the force free hard sphere model of relaxation~\cite{Hwang1975},
    the value of $\xi$ can in turn determine
  the local translational diffusivity, \Dlocal,
    of the biological water coupled to protein
    or soft matter surfaces
    or embedded in their interiors.
(Detailed equations for the calculation
    of both $s_{max}$ and \Dlocal
    are reviewed at the end of the theory section.)
With a series of differently positioned spin probes,
    one can map out the hydration dynamics
    at different sites of the macromolecule~\cite{Armstrong_jacs,Armstrong_apomb,McCarney2008,Cheng_Han_Lee_jmr,Ortony_njp,Kausik_pccp,Kausik2009a,Kausik_jacs}.

However, the analysis and experiment
    (\ie~eq.~\ref{eq:emax}) previously employed
    in these and other investigations
    expects only the saturation factor, $s(p)$,
    to vary with microwave power, $p$,
    and does not account for the fact that
    both the $T_1$ time of the spin labeled sample
    -- which is important for setting experimental parameters --
    as well as the leakage factor, $f$,
    vary with microwave power, as laid out in this work. 
Both these variations come predominantly
    from an underlying change in the
    $T_{1,0}$ time as the microwave power increases
\begin{align}
    T_{1,0}(p) \approx& T_{1,0}(0)
    +
    p
    \left. \frac{\partial T_{1,0}}{\partial p} \right|_{p=0}
    .
  \label{eq:lint10_der}
    \intertext{We denote this more compactly as}
    T_{1,0}(p) \approx& T_{1,0} + p \Delta T_{1,0}
    ,
    \label{eq:lint10}
\end{align}
where the parameter $T_{1,0} \equiv T_{1,0}(0)$
    quantifies the bulk relaxation time
    in the absence of microwave irradiation,
    while the new parameter $\Delta T_{1,0}$
    quantifies the variation
    of the bulk relaxation with 
    incident microwave power.
For the aqueous solution of
    free nitroxide spin probes
    studied here,
    $T_{1,0}(p)$ is exactly the relaxation time of pure water,
    which depends linearly on temperature,
    as discussed earlier.
Since both the heat capacity and
    the dielectric absorption coefficient
    (and therefore the conversion
    of microwave power into temperature)
    of water
    should remain relatively constant
    near ambient temperature,
    we can anticipate that eq.~\ref{eq:lint10}
    is a very good approximation.
Indeed, we have experimentally verified that $T_{1,0}(p)$
    remains linear with power
    (as will be shown in fig.~\ref{fig:leakage_diffconcs}).
One easily could,
    but typically will not need to, include any
    higher order terms, such as $p^2 \Delta^2 T_{1,0}$.
\subsubsection{Experimental Considerations}
In the past, researchers set the repetition delay
    between NMR signal acquisitions
    by assuming that the $T_1$
    of a sample remained constant,
    whether or not it was irradiated with microwaves.
They either acquired signal at $5\times$
    to $10\times T_1$
    or\footnote{Kausik and coworkers
        empirically arrived at the procedure of employing
        a value of $10\times T_1$ to obtain consistent ODNP data,
        without providing an analysis for this treatment.
    The analysis presented here accounts for the need
        for such a long delay.} acquired ODNP data
    with a fast repetition time  ($\approx 0.5\secs$)
    in an Ernst angle experiment.
We will demonstrate that,
    even though they represent the standard
    for most NMR experiments,
    the fast repetition experiments
    lead to large discrepancies
    in ODNP measurements.
Specifically, because heating increases the
    bulk water relaxation time $T_{1,0}$,
    it selectively suppresses the
    amount of signal detected
    at high microwave powers\footnote{Note
        that since the parameter $\Delta T_{1,0}$
        depends on the specific hardware,
        such as the details of the microwave cavity and ODNP probe,
        the power at which this effect sets in necessarily varies between
        different setups, including between different cavities.}
    ($p\gg T_{1,0}/\Delta T_{1,0}$), as shown below.
Because 
    this causes the $T_{1}$ of low concentration samples
    to lengthen considerably with
    even relatively small amounts of sample heating,
    this effect can easily lead to erroneous or inconsistent
    measurements of the enhancements;
however, with appropriate consideration,
    this situation is easily avoided.

We now analyze this signal suppression in detail.
As previously discussed,
    an ODNP experiment involves several scans
    that determine the enhancements
    at a series of different microwave powers, $E(p)$.
Each scan generally consists of
    a train of resonant rf pulses of
    a constant flip angle, $\theta$,
    separated by a constant repetition delay, $t_r$.
For instance, the optimized Ernst-angle
    experiment employs an angle such that
    $\cos(\theta)=\exp\left( -t_r/T_1 \right)$.
For a steady-state pulse train,
    the Bloch equations
  determine the fraction of
    available magnetization
    that the NMR relaxation between scans
 actually manages to recover;
this is
    \cite{Cavanagh_ernst}
    \newcommand{\thisexp}{\exp{\left( -t_r k_\rho C - t_r \left( T_{1,0} + p \Delta T_{1,0} \right)^{-1} \right)}}
    \begin{align}
        \frac{M_{pt}}{M_\infty}
        =&
        \frac{
        1-\exp\left( -\displaystyle\frac{t_r}{T_1} \right)
 }{
   1-\cos(\theta)\exp\left( -\displaystyle\frac{t_r}{T_1} \right)
        }
        \sin(\theta)  \label{eq:steady_state_unsubst}
        .
    \end{align}
Here, $M_{pt}$ gives the fraction of the ODNP-enhanced
    magnetization
    that the pulse train actually detects,
    while $M_\infty$ gives the total ODNP-enhanced
    magnetization
    in the absence of any pulses.

In order to explore how the change in $T_{1,0}$
    with power affects samples of
    different spin probe concentrations differently
    we find it useful to explicitly denote the concentration, $C$,
    by making use of the self-relaxivity constant, $k_\rho$,
    \begin{align}
 k_\rho = \frac{\rho}{C}
        \label{eq:relaxivity_def}
    \end{align}
    (table~\ref{tab:symbol_table}).
From eq.~\ref{eq:t1sum}
 (the relationship between $T_1$ and $T_{1,0}$),
    and eq.~\ref{eq:lint10} (the power dependence of $T_{1,0}$),
    we then retrieve the longitudinal relaxation
    for a given spin probe and hardware setup
    rate as a function of concentration and microwave power
    \begin{align}
        T_1^{-1}(p) =&
        k_\rho C + T_{1,0}^{-1}(p) 
            \notag\\
            =&
            k_\rho C + \left( T_{1,0} +p \Delta T_{1,0} \right)^{-1}
        .
        \label{eq:t1_vs_p_and_C}
    \end{align}
We can measure or estimate $\Delta T_{1,0}$
    for a specific hardware setup and
    type of sample composition,%
    \footnote{For practical experimental purposes,
            it can be illustrative to employ the definition
     $T_{1,0}^{-1}(p)
            =
            (T_{1,HH} + p\Delta T_{1,HH})^{-1}
            +
            (T_{1,w} + p\Delta T_{1,w})^{-1}$,
       where $T_{1,w}$ and $\Delta T_{1,w}$
            are the values for {\it pure water},
            and thus constant for a given experimental setup,
            while $T_{1,HH}^{-1}$ is the relaxation
            rate contribution
            of the \textit{unlabeled}
        sample, and which typically comes from
            proton-proton dipolar coupling,
            and is proportional to the concentration
            of the unlabeled sample.
            Typically, we can approximate
     $\Delta T_{1,HH}$ as 0.}
    and we can also roughly estimate
    the relaxivity, $k_\rho$.
Substitution of this value into eq.~\ref{eq:steady_state_unsubst},
    directly leads to the amount of
    signal suppression as a function of microwave power,
    \begin{align}
  \frac{M_{pt}}{M_\infty}
        =&
        \frac{\sin(\theta) \left( 1 -\thisexp \right)}{1-\cos(\theta)\thisexp}
        .
        \label{eq:steady_state}
    \end{align}
At high spin probe concentration,
    the power dependence remains minimal (since the $t_r k_\rho C$ term dominates),
    leading to equal signal suppression
    at all microwave powers.
Therefore, for instance,
    acquisition with a fast repetition delay
    at high concentrations
    gives accurate and reproducible data.
However, at low spin probe concentrations,
    the power-dependent term arising from the bulk relaxation
    (the second term inside the exponential in eq.~\ref{eq:steady_state})
    becomes more significant;
    therefore, the signal suppression ($M_{pt}/M_\infty$)
    varies with microwave power,
    as previously mentioned.
Such an experiment will not record 
    the {\it actual} ODNP enhancement, $E$,
    but rather,
    an apparent enhancement, $E M_{pt}/M_\infty$.
Because the signal suppression of eq.~\ref{eq:steady_state}
    obscures the true enhancement, we refer to it
    as an artifact.

To routinely avoid
    this artifact in the data,
    we simply employ eq.~\ref{eq:t1_vs_p_and_C}
    to predict the longest $T_1$
    that occurs during the ODNP experiment
    \begin{align}
        T_{1,max}
        =&
   \frac{T_{1,0}+ p_{max} \Delta T_{1,0}}{k_\rho C \left( T_{1,0} + p_{max} \Delta T_{1,0} \right) + 1}
        ,
        \label{eq:t1max}
    \end{align}
    which occurs for the ODNP scan that employs the
    maximum microwave power $p_{max}$.
Again, we remind the reader that all the parameters above
    are either known or can be estimated.
By employing a repetition time of at least $5\times T_{1,max}$,
 we can insure that
    the signal for each scan
    quantitatively includes all ODNP-enhanced magnetization,
    thus preventing artifacts.
Of course, to acquire reproducible 
    data, one must always measure
    the actual value of $T_{1,max}$ for a sample;
    this is not troublesome,
    since the same value
    is required for the correction
    described in the following section.
\subsubsection{Impact of Heating on ODNP Enhancements}
Now that the preceding section explains
    how to detect the actual enhancements, $E$,
    we can determine the effect of sample heating on
    the actual enhancements.
Specifically,
    we examine how the change in bulk water
    relaxation impacts the leakage factor,
    and therefore (via.~eq.~\ref{eq:polarization_subst})
    the enhancements.
One can combine eqs.~\ref{eq:feq},\ref{eq:lint10}, and \ref{eq:relaxivity_def}
    to quantify how the leakage factor, $f$,
    varies with microwave power, p, and concentration, C:
\begin{align}
    f(p,C) =&
    \frac{k_\rho C T_{1,0} \left( 1 +p \frac{\Delta T_{1,0}}{T_{1,0}} \right)}{1 + k_\rho C T_{1,0} \left( 1 +p \Delta \frac{T_{1,0}}{T_{1,0}} \right)}
    .
    \label{eq:f_linear}
\end{align}
Note that though the relaxivity, $k_{\rho}$,
    and thus the local water dynamics around the spin probes,
    does vary somewhat with temperature~\cite{Bennati_fcr}, we will find
    (in the results section)
    that for samples with concentrations of spin probe
    on the order of hundreds of micromolar or less
    (\ie~typically desirable concentrations for biological samples),
    the change in $p \Delta T_{1,0}$ overwhelms any variation
    due to changes in $k_{\rho}$.

Let us examine eq.~\ref{eq:f_linear}
    in the limiting extremes of
    spin probe concentration, $C$.
At high concentration,
    where the relaxation of the water protons near the spin probe
    dominates over the relaxation of the water protons in the bulk
    (\ie~$C k_\rho T_{1,0}\gg 1 $),
    $f(p)$ has a value close to 1
    and changes little with power.
On the other hand,
    at low concentration (\ie~$C k_\rho T_{1,0}\ll 1$),
    the denominator approaches 1 and $f(p)$
    varies linearly with power.

If one applies the uncorrected analysis (eq.~\ref{eq:emax})
    to enhancement data taken from low concentration samples,
    the change in the leakage factor
    with power expressed by eq.~\ref{eq:f_linear}
    will obscure the true value of the coupling
    factor, $\xi$.
From
    the expressions for
    $f(p, C)$
    (eq.~\ref{eq:f_linear})
    and
    $1-E(p)$
    (eq.~\ref{eq:polarization}),
    we can determine the effect of the changing
    leakage factor on the enhancements.
We take the ratio
    of the amount of polarization transferred ($1-E(p)$)
    in the presence and absence of
    microwave heating that changes
    the bulk relaxation time.
In other words, we can determine
    the ratio of the value expected by the
    uncorrected model to the actual value that we
    expect, including heating effects:
\begin{align}
    \frac{1-E_{\mbox{\scriptsize heating}}(p)}{1-E_{\mbox{\scriptsize no heating}}(p)} &=
    \frac{f(p)}{f(0)}\quad\quad\quad\quad\notag\\
    &=
    \frac{1 + k_\rho C T_{1,0}}{1 + k_\rho C T_{1,0} \left( 1+p\frac{\Delta T_{1,0}}{T_{1,0}} \right)} \left( 1+p\frac{\Delta T_{1,0}}{T_{1,0}} \right)
    \label{eq:f_is_linear}
\end{align}
As long as the fractional change in the bulk water relaxation,
    $p \Delta T_{1,0} / T_{1,0}$,
    remains small enough,
    this ``drift'' in the net polarization transferred
    remains roughly linear with microwave power.
We expect (and will see) that the drift
    in the leakage factor, $f$,
    leads to a linear variation of the enhancements
    in the high power regime (eq.~\ref{eq:f_is_linear}),
    while the previous theory (eq.~\ref{eq:emax})
    expects the enhancements to approach their asymptotic
    maximum and, therefore, remain constant.
This drift causes problems when attempting to reproducibly
    extrapolate the enhancements to
    the asymptotic maximum, $E_{max}$,
    as required by the uncorrected analysis.
Not only will it prevent data from fitting
    the model perfectly,
 but also, even slight changes to the characteristics
    of the hardware
    can lead to changes in $\Delta T_{1,0}$;
furthermore, changes to
    the range and spacing of the microwave powers, $p$,
    sampled by the experiment
    will lead to different weighting of
    data points with different leakage factors
    (eq.~\ref{eq:f_is_linear}).
Therefore,
    any attempt to extrapolate to
    the asymptotic maximum, \Emax, will necessarily
    give varying results.
As a result,
    for samples with nitroxide probe concentrations
    below about 1\mM ($\approx 1 / k_\rho T_{1,0}$ for bulk water),
    the uncorrected analysis (eq.~\ref{eq:emax})
    only gives approximate values
    for the coupling factor, $\xi$.
Still, the uncorrected analysis can accurately identify
    meaningful changes in the coupling factor
    ($\xi$) value
    and thus identify changes or trends in the
    translational hydration dynamics
    for similar samples.
However, the reproducibly of such measurements
    critically depends on the exact
    duplication of both the experimental and hardware
    parameters for all the $\xi$ values that are
    being compared.
\subsubsection{The Corrected Analysis}\label{sec:corrected_analysis_w_outline}
Having reviewed the relevant parts of the existing ODNP theory
    and pointed out its limitations,
    we now seek to increase the accuracy with
    which we extract the dynamic parameter
    $\xi$ from the enhancement values.
We seek a new approach to data acquisition and
    analysis that can correct
    for those errors
    that arise as a result of dielectric absorption and heating
    (\ie~eq.~\ref{eq:f_is_linear}).
At the same time,
    we seek a method that can extract meaningful
    information about the hydration dynamics
    at low concentrations,
    where the leakage factor ($f$)
    (determined in eq.~\ref{eq:feq_exp}
    and used in eq.~\ref{eq:emax})
  approaches zero,
    and thus would appear to make the
    coupling factor ($\xi$) ill-determined.

The equation that gives the enhancements (eq.~\ref{eq:polarization})
    has historically been
    phrased in terms of the unitless leakage factor, $f$,
    and the unitless coupling factor, $\xi$.
As we can already see from the complexity of eq.~\ref{eq:f_is_linear},
    this strategy poses a problem when attempting to analyze
    concentration-dependent effects such as dielectric heating.
However, Hausser and Stehlik~\cite{Hausser1968} derive
    $f$ and $\xi$ from three more fundamental rates:
    the rate of local dipolar
    cross-relaxation
    between the electron and the proton, $\sigma$,
    the local dipolar
    self-relaxation rate of the protons, $\rho$,
    and
    the intrinsic relaxation rate of the bulk water protons, $T_{1,0}^{-1}$
    (all of which have units s$^{-1}$),
which includes any
  relaxation driven by mechanisms
    that do not involve the spin probe.
To these parameters, we add
    $\Delta T_{1,0}$, which is determined
    by the particular hardware configuration.

We note that the $T_{1,0}$ time
    depends only on the
    characteristics of the
    unlabeled solution.
By contrast,
    $\sigma$ and $\rho$ come from interactions
    that scale with distance, $r$, from the spin
    label as $r^{-6}$.
Thus, they depend only on the local water dynamics
    in the sample under investigation,
    and scale with spin label concentration.
Therefore, rather than referring to $\rho$ and $\sigma$,
    we can refer to the
    self-relaxivity, $k_\rho$,
    defined earlier (eq.~\ref{eq:relaxivity_def})
    and the similarly defined cross-relaxivity
    \begin{equation}
        k_\sigma = \frac{\sigma}{C}
        .
        \label{eq:ksigma_def}
 \end{equation}
\newcommand{\relaxivityunits}{\xspace\ensuremath{\mbox{s}^{-1}\mbox{M}^{-1}}\xspace}%
These relaxivity parameters have units \relaxivityunits
  (table~\ref{tab:symbol_table}).

The corrected analysis now separately determines
    $k_\sigma$
    (from the measurements of $E(p)$ and $T1(p)$)
    and
    $k_\rho$
    (from the measurements of $T_{1,0}$ and $T_1$),
    as will be detailed below. 
While the values $E_{max}$ and $f$
    of the uncorrected analysis (eq.~\ref{eq:emax})
    are both dependent on --
    but not linearly proportional to --
    the spin probe concentration,
    the bulk water relaxation rate, $T_{1,0}^{-1}$,
    and a heating-induced term
    of the form
    $p \Delta T_{1,0}$ (from eq.~\ref{eq:lint10}),
    the values of $k_\sigma$ and $k_\rho$ 
    depend on neither the spin probe concentration,
    nor $T_{1,0}^{-1}$, nor $\Delta T_{1,0}$.
As we will demonstrate,
    this allows one to bypass the heating effects
    and to determine meaningful information
    at low concentration.
The ratio of these two local relaxivities
    \begin{align}
        \xi = \frac{k_\sigma}{k_\rho}
        \label{eq:coupling_factor}
        ,
    \end{align}
    gives the unitless coupling factor,
 and thus information about the
    translational correlation time
    $\tau_c$, which translates to a
    unique value for the local translational diffusivity,
    $D_{local}$.

We now outline the specifics of the analysis
    described above,
    which consists of the following steps:
\begin{enumerate}
    \item Interpolate a small set of measurements\footnote{For the measurements in the results section,
        we typically measure $T_1(p)$ in the absence of microwave power, at $p_{max}$ and $p_{max}/2$.
    Occasionally, we acquire the powers in 5 steps.}
        of $T_1(p)$
          to generate values corresponding to all microwave powers at which
          we measure the ODNP signal enhancements, $E(p)$. 
    \item Determine a set of $k_\sigma s(p)$ values at all
        microwave powers, $p$.
    \item Find the asymptotic limit, $k_\sigma s_{max}$.
    \item Determine $k_\sigma$.
    \item Perform $T_1$ measurements in the absence of microwave power,
        on both the spin labeled sample and an unlabeled reference
        sample, in order to determine $k_\rho$. 
    \item Determine $\xi = k_\sigma/k_\rho$ which in turn
        yields the $\tau_c\propto \left( D_I+D_{SL} \right)^{-1}$ value.
\end{enumerate}
\renewcommand{\paragraph}[1]{

\quad

\hspace{-0.75cm}\textbf{#1)}

}
\paragraph{Step 1}
In order to quantify the $k_{\sigma}s_{max}$ values
    to ultimately determine $k_\sigma$,
    one must first determine
    $T_1(p)$ values for each $E(p)$ data point.
Specifically, these values will be employed
    in step 2 to correct for heating
    effects.
Typically, one does not have the experimental
    time to measure the $T_1(p)$ at all microwave powers ($p$),
    and so we need a method for interpolating
    the relatively few $T_1(p)$ values
    we measure to reasonably predict $T_1(p)$ at
    all microwave powers.
From eq.~\ref{eq:t1_vs_p_and_C},
    we recall
    the NMR $T_1$ rate in the absence of microwaves:
  \begin{equation}
      T_1^{-1}(0) = k_\rho C +  T_{1,0}^{-1}
      \notag
        \label{eq:F_linear_der1}
 \end{equation}
    as well as the rate we expect 
    when microwave power is applied
    (\ie eq.~\ref{eq:t1_vs_p_and_C} identically):
\begin{equation}
        T_1^{-1}(p) = k_\rho C + \left( T_{1,0} + p \Delta T_{1,0} \right)^{-1}
       .
        \label{eq:F_linear_der2}
  \end{equation}
Subtraction of these two equations,
    followed by some rearrangement leads to the value\footnote{we
    note that the left side of the approximation
     still varies close to linearly w.r.t. microwave
    power, even for samples where the approximation
     does not hold in the absolute sense}
 \begin{equation}
  F_{linear}(p)
  \equiv
        \frac{1}{T_1^{-1}(p)-k_\rho C}
        \approx 
        T_{1,0} + p \Delta T_{1,0}
        \label{eq:F_linear}
        .
    \end{equation}
We can determine $k_\rho C$ from the $T_1$
    data with power off (eq.~\ref{eq:F_linear_der1}),
    namely
    \begin{equation}
        k_\rho C = T_1^{-1}(0) - T_{1,0}^{-1}
        \label{eq:rho_at_power_off}
        .
    \end{equation}
As noted in eq.~\ref{eq:F_linear},
    $F_{linear}$ is an approximately linear
    function of microwave power.
Thus, by fitting the values of
    $F_{linear}(p)$
    to a straight line, then solving for
    \begin{equation}
        T_1(p)
        =
        \frac{1}{F_{linear}^{-1}(p)+k_\rho C}
        \label{eq:t1_from_F_linear}
    \end{equation}
    one can retrieve an accurately interpolated
    value for $T_1(p)$.

We make two practical notes at this point:
First,
    a thresholding procedure
    at high concentration
    where $T_1(p)^{-1}\approx k_\rho C$
    is required to
 manually set $T_1(p)$ to $k_\rho C$
    in order to
    prevent numerical blow-up of $F_{linear}$. 
Second, this interpolation procedure can be
    performed without the need for a $T_{1,0}$ measurement.
Eq.~\ref{eq:F_linear}
    and eq.~\ref{eq:t1_from_F_linear} do not depend
    very sensitively
    on the value of $T_{1,0}$.
A reasonable first estimate of $T_{1,0}$
    will usually suffice,
    and when it does not, one can determine
    $T_1(p)$ from a very closely spaced interpolation.
\paragraph{Step 2}
With the help of
    eq.~\ref{eq:coupling_factor}, \ref{eq:feq}, and \ref{eq:relaxivity_def},
    we rewrite the previous equation for
    the signal enhancements (eq.~\ref{eq:polarization})
    in terms of the fundamental relaxivities
    \begin{equation}
        1-E(p) =
        \obpair{\frac{k_\sigma}{k_\rho}}{$\xi$}
        \obpair{\frac{k_\rho C}{k_\rho C + T_{1,0}^{-1}(p)}}{$f$}
       \left| \frac{\omega_e}{\omega_H} \right|
       s(p)
       .
    \end{equation}
We can now multiply by
    the form of $T_1(p)$ given by
    eq.~\ref{eq:t1_vs_p_and_C} ($T_1(p) = 1/(k_\rho C+T_{1,0}^{-1}(p))$),
    then cancel and rearrange terms to arrive at
    \begin{equation}
        k_\sigma s(p)
        =
        \frac{1-E(p)}{C T_1(p)}
        \left| \frac{\omega_H}{\omega_e} \right|
        .
        \label{eq:polarization_subst_fundamental0}
    \end{equation}
Note that this is mathematically equivalent to
    the typical equation for the enhancements,
    eq.~\ref{eq:polarization}.
However,
    it clearly illustrates how the
    variation of the $T_1(p)$ time with
    microwave power,
    as given by eq.~\ref{eq:F_linear_der2} (\ie~eq.~\ref{eq:t1_vs_p_and_C}),
    perturbs the DNP signal enhancements.
By inserting a value 
    of $T_1(p)$ that is experimentally measured at
    (or interpolated to)
    the correct microwave power,
    one can fully account for the
    change in $T_{1,0}(p)$
    that arises from dielectric heating.
Inserting the interpolated $T_1(p)$
    values from step 1 into
    eq.~\ref{eq:polarization_subst_fundamental0}
    should yield $k_\sigma s(p)$
    values that
    depend asymptotically on power
    \begin{eqnarray}
        \frac{1-E(p)}{C T_1(p)}
        \left| \frac{\omega_H}{\omega_e} \right|
        &=&
        k_\sigma s(p)
        \notag
        \\
        &=&
        \frac{k_\sigma s_{max} p}{\phalf+p}
        \label{eq:polarization_subst_fundamental}
    \end{eqnarray}
    (where we have substituted eq.~\ref{eq:smax}),
    in spite of mismatch of
    the enhancements, $E(p)$,
    to the uncorrected model
    (\ie eq.~\ref{eq:polarization},\ref{eq:emax}).
Thus, one can then proceed to extract
    accurate values of $k_\sigma s_{max}$
    in spite of the drift in the enhancements
    at high microwave powers
    that we previously noted in eq.~\ref{eq:f_is_linear}. 
\paragraph{Step 3}
Extrapolating eq.~\ref{eq:polarization_subst_fundamental}
    to infinite power then
    yields $k_\sigma s_{max}$
    in the same fashion that the uncorrected
    analysis extrapolated the signal
    enhancements to infinite power to find $E_{max}$, \ie
\begin{equation}
        k_\sigma s_{smax}
        =
        \lim_{p\rightarrow\infty}
      \left( 
        \frac{1-E(p)}{C T_1(p)}
 \left| \frac{\omega_H}{\omega_e} \right|
       \right)
        .
        \label{eq:ksmax_extrap}
    \end{equation}

Note that the previously employed, uncorrected, analysis
    employs measurements of both $f$ or $\xi$.
In order to determine either of these values,
    one must perform
    measurements on two samples:
    both the spin-labeled
    sample and the sample where the spin probe has
    been removed.
Therefore, sample preparation issues with {\it either}
 sample can lead to errors in the measurement,
    as can discrepancies between the two samples.
By contrast, note how one can determine $k_\sigma$
    from measurements on {\it only} the spin-labeled sample.
\paragraph{Step 4}
As with the uncorrected model,
    Armstrong's model~\cite{Armstrong_jcp} then predicts
    a value for $s_{max} \approx 1$
    for most spin labeled biological or soft matter samples,
    giving $k_\sigma \approx k_\sigma s_{max}$ directly.
In the case of samples with freely dissolved spin label,
    as employed in Bennati \etal~\cite{Turke_sat},
    we can neglect the effect of nitrogen relaxation.
We can then calculate the value of
    \begin{equation}
        s_{max} = 1 -
        \frac{2}{3+3 b''}
        \label{eq:hyde_smax}
    \end{equation}
    for $^{14}N$ nitroxides
    where,
    following the work of Hyde and Freed~\cite{HydeFreed},\footnote{To 
        come to this conclusion,
        we also incorporate the deduction that (for $^{14}$N)
        $s_{max} = \frac{1}{3}-\frac{2}{3}R$,
        where $R$ is the ``reduction
        factor'' from \cite{HydeFreed}.}
    $b'' = w_h/6 w_e$
    gives a ratio between
    the rate of Heisenberg exchange, $w_h$, and the
    rate of electron relaxation,  $w_e$.
ELDOR (electron double resonance) curves
    that measure both numbers on $^{15}$N nitroxides
    are presented by Bennati~\etal in~\cite{Turke_sat}
    and lead to a value of $b'' = C \times 198.7\M^{-1}$,
    where $C$ is the spin label concentration.
\paragraph{Step 5}
By referencing the relaxation time of the spin
    labeled sample to that of the unlabeled sample,
    we can determine $k_\rho$.
Specifically,
  \begin{equation}
       k_\rho = \frac{T_1^{-1}-T_{1,0}^{-1}}{C}
        \label{eq:k_rho_eq}
        .
    \end{equation}
We determine $k_\rho$ entirely from measurements
    in the absence of microwave power,
    so that,
    as can be seen from eq.~\ref{eq:F_linear_der2},
    dielectric heating effects are not an issue.
It is also clear from eq.~\ref{eq:k_rho_eq} that
    the determination of $k_\rho$
    (as well as $k_\sigma$ cf.~eq.~\ref{eq:ksmax_extrap})
    requires an accurate
    knowledge of the spin label concentration, $C$.
One can often determine accurate concentrations
    of both spin labeled and non spin labeled biomolecules via UV-visible
    spectrophotometry
    or determine accurate concentrations of spin labeled
    biomolecules
    via ESR.
However, as will be discussed in the next step,
    the accurate determination of the coupling factor
    -- the ultimate parameter of interest --
    again does not require
    knowledge of the absolute concentration.
\paragraph{Step 6}
Finally, we find the ratio of the two relaxivities
    (\ie $k_\sigma$ and $k_\rho$)
    to give the coupling factor, $\xi$
    (\ie eq.~\ref{eq:coupling_factor}),
    and -- by extension -- the local translational diffusivity
    of the hydration water.
Up to this point,
    it is true that the individually determined
    relaxivities
    are susceptible to systematic errors in the actual concentration
    of spin probe, $C$, relative to the nominal concentration of
    the spin probe (see expressions for the relaxivities in 
    $k_\sigma$ cf.~eq.~\ref{eq:ksmax_extrap}).
However, issues with solubility, dilution, etc.  can scale
  the concentration in both the labeled and unlabeled
    sample.
During calculation and application of the leakage factor, $f$,
    in the previous analysis, such errors could be factored
    out,
    as long as samples were prepared in the same way.
This final step of the newly proposed analysis
    also cancels such systematic
    errors,
    \ie $\xi = ( k_\sigma \cancel{C} )/( k_\rho \cancel{C} )$,
    (eq.\ref{eq:coupling_factor})
    in the same fashion.
\newcommand{\xbulk}{\ensuremath{\xi_{bulk}}\xspace}%
\newcommand{\xsite}{\ensuremath{\xi_{site}}\xspace}%
\newcommand{\tausite}{\ensuremath{\tau_{c,site}}\xspace}%
\newcommand{\Dw}{\ensuremath{D_{H_2O}}\xspace}%
\newcommand{\Dsl}{\ensuremath{D_{SL}}\xspace}%
\newcommand{\taubulk}{\ensuremath{\tau_{c,bulk}}\xspace}%

The local translational diffusion coefficients
    can then be determined from
\begin{equation}
    \Dlocal
    \approx
    \left( \Dw+\Dsl \right)
    \left( \frac{\taubulk}{\tausite} \right)
    .
    \label{eq:calcD}
\end{equation}
Here, $D_{SL}=4.1\magn{-10}~\mbox{m$^2$s$^{-1}$}$
    (the approximate diffusivity of small TEMPO derivatives)
    and
    $D_{\mbox{H$_2$0}}=2.3\magn{-9}~\mbox{m$^2$s$^{-1}$}$~\cite{Armstrong_jacs}.
Each \correltime is the
    translational correlation time of
    a spectral density function
    that predicts the observed coupling factor
    (\xbulk or \xsite).
Physically, this correlation time can be described as the
    lifetime of the dipolar interaction between the electron spin
    of the spin probe and the proton spin of the water molecule.
We note that the
    ``translational correlation time'' is not necessarily a
    uniquely defined property of the molecular dynamics of a
    particular system.
Rather, its exact value will also depend on the nature of the
    interaction probed by a particular measurement.
Thus, we can expect that,
    even though they should exhibit similar trends and relative
    values,
    the translational correlation times
    generated by dynamics stokes shift
    spectroscopy~\cite{Zhong_cpl},
    ODNP, as well as various scattering
    measurements might well be different from each other,
    even for systems with the same dynamics.
By analogy,
    the rotational correlation time differs for different measurements,
    depending on whether the measurement probes the relaxation
    of an interaction that depends on rank one spherical harmonics
    (e.g. dielectric spectroscopy)
    or rank two spherical harmonics
    (e.g. NMR quadrupolar relaxation)~\cite{Bottcher1974,Huntress1970,Hindman1973,oleinikova2004can}.
This highlights the usefulness of translating this number
    (even approximately) to a local translational diffusion,
    \Dlocal, which \textit{is} uniquely defined based solely
    on the local molecular dynamics; 
    this is precisely the
    role of eq.~\ref{eq:calcD}.

The coupling factor,
    \xbulk,
    corresponding to
    the correlation time of unrestricted
    water, \taubulk,
    comes from the analysis
  of samples of small,
    freely dissolved spin-probes;
in the results section,
    we will confirm a result of
    $\xbulk = \xireferenceval$.
The value of the coupling factor is specific
    to the approximate static field, $B_0$,
    employed in the ODNP experiment.
Following previous literature \cite{Bennati_fcr,Turke2010,Armstrong_jcp,Armstrong_jacs,Armstrong_apomb,HodgesBryant}
    we employ Hwang and Freed's
    expression for the spectral density function
    of the dipolar interaction between the spin probe
    and the water.
This expression derives from a force free hard sphere (FFHS)
    model of translational diffusion and is proportional to
    \begin{equation}
 J(z)
        =
        \Real{ \frac{1+\frac{z}{4}}{1+z+\frac{4 z^2}{9}+\frac{z^3}{9}} }
    \end{equation}
    for which 
 \begin{equation}
      z = \sqrt{i \omega \tau_c}
        \label{eq:define_Freed_z}
    \end{equation}
    and it determines the
    functional form of the spectral density function,
    \begin{eqnarray}
        &&\xi(B_0,\tau_c)=\\
        \nonumber
        &&\quad
        \frac{6 J\left( (\gamma_e-\gamma_H) B_0 \tau_c \right)
            - J \left( (\gamma_e+\gamma_H) B_0 \tau_c \right)
        }{
            6 J\left( (\gamma_e-\gamma_H) B_0 \tau_c \right)
            + 3 J\left( \gamma_H B_0 \tau_c \right)
   + J\left( (\gamma_e+\gamma_H) B_0 \tau_c \right)
        }
  ,
        \label{eq:xi_fn_of_J}
 \end{eqnarray}
    where $\gamma_e=g \mu_B/h$ and $\gamma_H$
    are the gyromagnetic ratios for the electron
    and proton spin, respectively.\footnote{Note also that
    eq.~\ref{eq:xi_fn_of_J} always applies to dipolar
    interactions, regardless of the particular choice of the
    spectral density function, $J(\omega)$.} %
In order to generate this coupling factor, \xbulk, of \xireferenceval
    at a static field
    ($B_0 = \omega_H/\gamma_H$)
  corresponding\footnote{Here, the gyromagnetic ratio for protons is
      $\gamma_H = 2 \pi \times 4.258\magn{7}\;\left[ \mbox{Hz} \mbox{T} \right]$}
    to the experimental NMR resonance frequency of $\omega_H/2\pi = 14.8\MHz$
    we must choose an FFHS spectral density function with \taubulkequation.
Note that eq.~\ref{eq:calcD}-\ref{eq:xi_fn_of_J}
    apply equally well to the coupling factor
    determined either from the corrected analysis,
    as presented here,
    or from a coupling factor
    that the previously employed, uncorrected, analysis
    determines from leakage
    factor, $f$, and asymptotic enhancement, $E_{max}$ values.
\section{Results}
\outlineblank{subsection}
\subsection{Heating Artifacts}
\begin{figure}[tbp]
    \begin{center}
        \includegraphics[width = 3.5in]{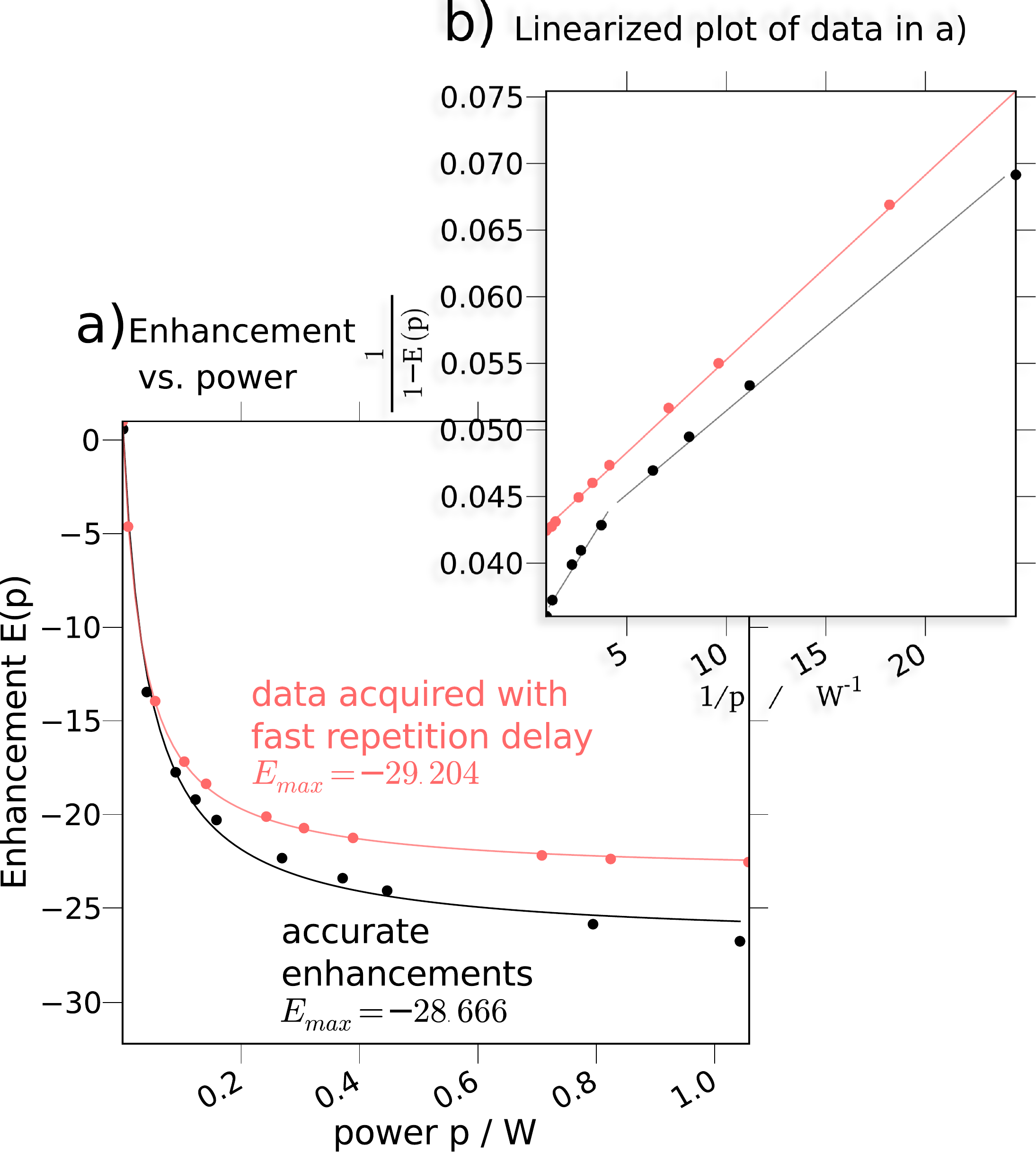}
    \end{center}
    \caption{Fast repetition delays
        can generate a misleading
        artifact that alters the apparent enhancement
        as a function of microwave power.
    This artifact
        (here the difference between the red (gray) and black curves)
        can vary depending on the experimental setup
        and the exact value chosen for the repetition delay.
  It also obscures the underlying physics at work.
    We present data acquired with a 0.5\secs repetition delay
        whose parameters and results (in red (gray))
 match the parameters and results in \cite{Armstrong_jcp}.
    An initial analysis would lead one to interpret the fact that
        the apparent enhancements (see eq.~\ref{eq:steady_state}),
        $EM_{pt}/M_\infty$ -- shown in red (gray),
        level off at high microwave
        powers (here, at powers higher than 0.4\W)
        as fairly rigorous evidence that the
        ESR transition saturates at those powers,
        following eq.~\ref{eq:polarization_subst}.
    However,
        we repeat the experiment
    with a repetition delay
        exceeding $5 \times$ the $T_1$ at maximum power
        (\ie~$T_{1,max}$, eq.~\ref{eq:t1max}),
        but otherwise identical conditions,
        and see that the amount of polarization transferred
        at high microwave power actually continues to increase
        in an approximately linear fashion.
    This latter set of data (in black) recovers all of the magnetization, and therefore
        more {\it accurately} quantifies the amount of polarization transferred,
        $1-E$, by ODNP,
        even thought it
        violates our previous expectations
        in that the enhancements, $E$, do not
        depend asymptotically on power
        as predicted by eq.~\ref{eq:polarization_subst}.
    The best-fit asymptotic curve
        (in black) emphasizes this mismatch.
    Similarly, as seen from the inverse of
        eq.~\ref{eq:polarization_subst},
        a plot of $1/(1-E)$ vs.~$1/p$
        indicates an apparent match to
        the uncorrected model
        when it presents a straight line,
        as it does for the apparent
        enhancements of the fast repetition delay experiment
        in red (gray)
        and highlights the mismatch to the uncorrected
        model for the actual enhancements (in black)
        by presenting two regions with different slopes
        (at powers higher and lower than approximately 5$\W^{-1}$).
    We note that the \Emax fits presented here
        are only presented to give a comparison 
        of the two curves.}
    \label{fig:Emax_mismatch}
\end{figure}

Among other results,
    the data by Armstrong \etal in~\cite{Armstrong_jcp}
    provide one of the earlier benchmarks
    for ODNP measurements of the coupling factor
    between bulk water and small spin probes.
The enhancement vs.~microwave power,
    $E(p)$, experiments in~\cite{Armstrong_jcp}
    are set up as an Ernst-angle
    NMR signal acquisition
    with a repetition delay of 0.5\secs.
Repeating this experiment
    with the same parameters,
    we find that the apparent enhancements,
    $E M_{pt}/M_\infty$
    (see eq.~\ref{eq:steady_state}),
    clearly level off at high power
    and apparently approach an asymptote,
    as presented in fig.~\ref{fig:Emax_mismatch}.
Different coupling factor
    values reported by other researchers~\cite{Bennati_fcr,Turke_sat,Sezer_xi,Turke2010},
    inconsistencies in our own repeated measurements of the coupling factor,
    and the theoretical analysis just presented (eq.~\ref{eq:steady_state})
    all motivate us to suspect the accuracy of these
    apparent enhancement values, especially since $0.5\secs \ll T_{1,max}$.

By acquiring a second set of data under the same
    conditions,
    but with a repetition delay that exceeds $5\times T_{1,max}$,
    we retrieve the accurate enhancement values,
    $E$ (fig.~\ref{fig:Emax_mismatch}),
    for this sample and setup.
As expected, increasing the relaxation delay
    beyond $5\times T_{1,max}$
    does not significantly affect the results
    for the enhancements, $E(p)$.
We observe that the actual amount of polarization
    transfered ($1-E$) at the highest microwave powers
    employed here (fig.~\ref{fig:Emax_mismatch})
    does not approach an asymptote
    but, rather, continues
    to increase as an approximately linear 
    function of microwave power.
Thus,
    the apparent enhancements, $E M_{pt}/M_\infty$,
    obtained with a 0.5\secs repetition delay experiment
    level off with high microwave powers,
    not as a result of
    saturation of the ESR transition
    but, rather, as a result of the artifact
    of employing increasingly insufficiently long repetition
    delays leading to artificially suppressed enhancement values,
    $E$, as described by eq.~\ref{eq:steady_state}. 

We remind the reader that
    the magnitude of the artifact identified here
    will vary with the particular hardware setup,
    and even the exact positioning of
    the sample within the same microwave cavity
    and setup.
For the remainder of the data presented here,
    we always ensure
    that the repetition delay exceeds $5\times T_{1,max}$.

One can now see how the determination of the actual
    enhancements, $E$, can present a
    significant roadblock for the previous analysis
    (eq.~\ref{eq:polarization_subst}).
A significant residual remains
    after the enhancements are
    fit to the asymptotic
    functional form that one 
    previously expected (\ie  eq.~\ref{eq:polarization_subst}).
This prevents one from accurately extrapolating
    to the asymptotic limit, $E_{max}$. 
As presented in the following sections
    (\ref{sec:rel_times}-\ref{sec:application_of_anal}),
    one must invoke the corrected analysis
    presented here
    (eq.~\ref{eq:polarization_subst_fundamental},\ref{eq:ksmax_extrap})
    in order to account for the slope
    in the enhancements at high microwave powers,
    and to accurately extract the 
    hydration dynamics values from such data.
\subsection{Identifying Sample Heating in ODNP Probes}
\begin{figure*}[tbp]
        \centering
        \includegraphics[width = 7in]{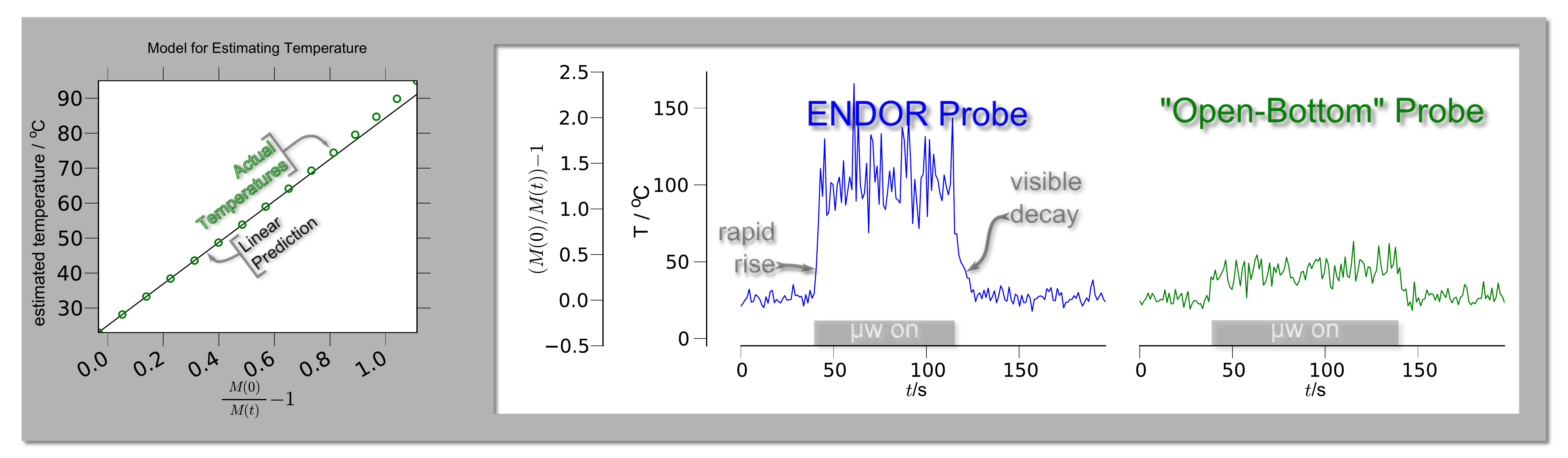}
        \caption{The simulated data in the left pane demonstrates 
            how the short saturation-recovery $T_1$ experiment
            can predict the temperature of the sample.
        The value $\left( M(t)/M(0) \right) -1$ is
            the relative difference in the observed signal, $M(t)$.
        This plot assumes an initial temperature
            of $\sim25^o$C,
            and thus an assumed initial $T_{1,0}$
            of $\sim2.5\secs$.
        The right pane displays the
            time-dependent response of the heating
            while cycling the microwave power from off to on.
        Data are presented
            both for the previously described home-built
            (``open-bottom'') probe design
            \cite{Armstrong_jcp,Armstrong_jacs,Armstrong_portable},
            as well as a sample positioned inside an ENDOR
            cavity.
        We find that the relatively ad-hoc, home-built
            ``open-bottom'' probe design
            demonstrates less sample heating
            than the commercial ENDOR setup.
        In addition, the time-dependent measurement
            of temperature in the ENDOR setup shows
            a decay that stretches over a few seconds,
            while this experiment cannot resolve the timescale
            of the temperature decay in the ad-hoc probe.
            }
    \label{fig:Tvst}
\end{figure*}
As discussed earlier, the measurements of $T_{1,0}$
    can probe both the magnitude and the timescale of the
    change in sample temperature that microwave irradiation
    induces.
We present two separate experiments,
    optimized, respectively,
    for measuring how the sample temperature varies
    as a function of the steady-state microwave power,
    as well as in time after a rapid change in microwave power. 

A numerical calculation
    (left pane of fig.~\ref{fig:Tvst}),
    based on
    the exponential $T_1$ decay
    (eq.~\ref{eq:t1fit})
    and the Hindman model
    (eq.~\ref{eq:wood_rotational})
    verifies that
    the relative change
    in magnetization,
    $\left(M(0)/M(T)\right)-1$, 
    remains a linear function of temperature
    from 20$^o$C to 80$^o$C during
    the single-scan short-recovery $T_{1,0}$ experiments
    previously described in sec.~\ref{sec:intrinsic_probe}.
This linear relationship relies on 
    both the approximately linear dependence of magnetization
    on $T_{1,0}$ when short recovery delays are
    employed (here $\tau = 0.5\secs \ll T_{1,0}$)
    and on the approximately linear dependence of $T_{1,0}$ on temperature.

These single-scan short-recovery $T_{1,0}$ experiments can
    quickly test the performance of the NMR probe configurations
    best suitable for ODNP experiments.
Specifically,
    since a commercially designed ENDOR cavity
    has a built-in rf coil,
    we expected that it might present less
    perturbation of the electric field in the cavity,
    and so demonstrate
    less sample heating
    or a significantly faster temperature response
    than home-built ODNP probe designs.
We performed the single-scan short-recovery $T_{1,0}$
    experiment on water loaded into a capillary
    and placed inside a commercial ENDOR cavity (Bruker~ER~801).
In this setup,
    variable capacitors were
    added inside an aluminum box
    outside the cavity in order to
    tune and match the ENDOR rf coil
    so it could be used for NMR signal detection.
In order to position the sample
    in the center of the cavity,
    a slightly larger capillary (1.2\mm o.d.)
    with one fused end
    holds the sample capillary (\capillarydim). 
We then compared the performance of this
    ENDOR probe to the home-built
    NMR saddle-coil probe design previously described
    in~\cite{Armstrong_jcp,Armstrong_portable,Pavlova_pccp,Cheng_Han_Lee_jmr}.
After loading a sample
    capillary into a homebuilt NMR probe,
    we inserted it into a typical TE101
    rectangular microwave cavity
    and again performed the
    single-scan short-recovery $T_{1,0}$ experiment.
Surprisingly,
    the temperature of the sample
    in the home-built ODNP probe
    remains lower overall and
    responds on a similar or faster timescale
    than the temperature of the sample
    in the ENDOR cavity
    (fig.~\ref{fig:Tvst}),
    \ie presents better ODNP performance. 
This data also offers insight into the
    time that these ODNP probes need to
    equilibrate the temperature after
    the application of microwave irradiation.
For both the ENDOR and the home-built setups,
    the temperature
    of the sample responds to changes in the
    microwave power within less than five seconds.
Thus, we can acquire accurate ODNP enhancement
    values without including additional waiting periods
    longer than the length of
    the $5\times T_1$ recovery
    period, \ie typically no more than 10-12\secs. 

Like others (e.g.~\cite{Turke2010}),
    we initially assumed that
    the transfer of heat through the capillary wall does not
    likely vary significantly between
    different probes and hardware setups.
However, the unexpected difference
    in sample heating observed in the ENDOR vs.~home-built probe
    indicates that the difference in their heat transfer
    can be significant.
Specifically, in the ENDOR setup, two layers of quartz
    (the sample capillary wall
    and the wall of the outer capillary used for positioning)
    and an intermediate, insulating layer of air
    come between the cooling air and the sample.
We hypothesize that this insulation may cause
    increased heat retention in the sample.
This would imply that the transfer of
    heat through the capillary wall
    contributes significantly towards cooling the sample
    and should be considered when designing an ODNP probe.

With this knowledge,
    we present an ODNP probe
    in which the NMR probe
    and sample are contained inside a
    3\mm quartz tube
    that passes entirely through the top and bottom openings
    of the microwave cavity (``pass-through design'').\footnote{More technical
        specifications for this probe will be introduced in an upcoming
        publication:~\cite{FranckHardware}.}
In the previously described home built design (\ie ``open-bottom design''),
    the sample tube and rf coil protrude from
    an enclosed glass tube that is held
    in the cavity from the top,
    and where the cooling air enters
    the cavity through the microwave waveguide
    on the side of the cavity.
Therefore, small variations in the iris coupling\footnote{\ie~movement of the dielectric
    insert that varies the coupling of the microwave into the cavity.}
    and sample positioning can
    lead to variation in
    the air-flow near the sample. 
In the new pass-through design,
    air flows through the 3\mm quartz tube,
    and thus over and across the NMR coil and sample capillary
    (\capillarydim, for both designs),
    resulting in more consistent
    sample cooling.
Additionally, in the previously described ``open-bottom'' design,
    only the collet at the top of the cavity
    stabilizes the sample position,
while the pass-through design
    holds the sample tube more firmly and reproducibly
    at the center of the cavity by fastening at the top and bottom,
    preventing radial displacement
    of the sample capillary.

\begin{figure}[tb]
    \begin{center}
        \includegraphics[width=3.5in]{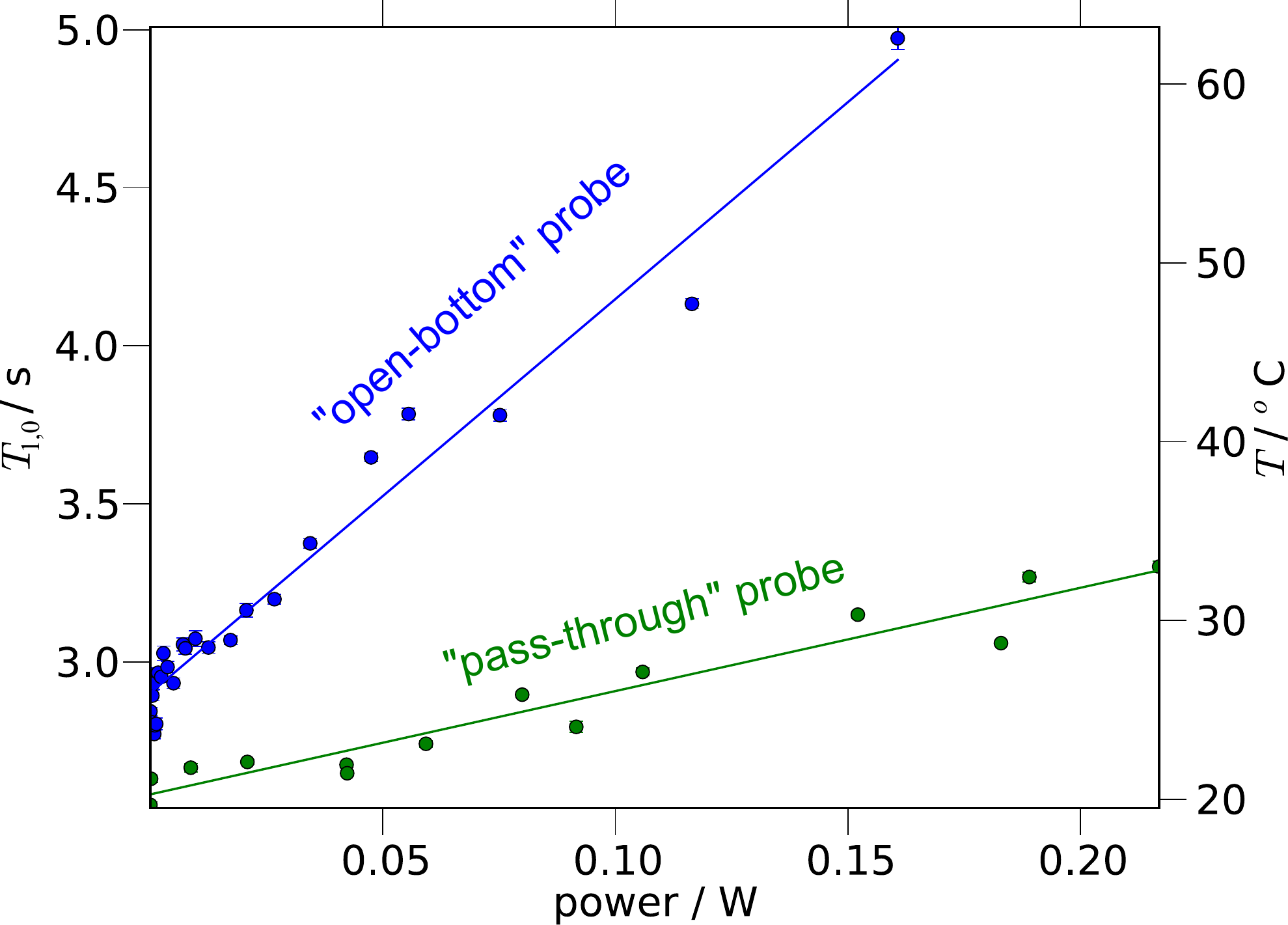}
    \end{center}
    \caption{A series of experiments show how the \tonen\ of pure water increases with increasing microwave power for two different hardware setups.
    The model of fig.~\ref{fig:hindmantemp} (from \cite{Hindman1973})
        then converts the $T_1$ values to temperatures  ($y-$axis on the right).
    Even with cooling air,
        the temperature can increase significantly with power
        in the open-bottom setup
        (design presented in
            previous publications~\cite{Armstrong_jcp,Armstrong_jacs,Armstrong_portable})
        and the curve presented here
        can vary dramatically as the probe position
        or the cavity matching changes.
    However, the improved pass-through probe design \cite{FranckHardware}
        reduces the heating to a reproducible level
        that is less than the lowest heating seen with the
        open-bottom design,
        while retaining the same sample size of 3.5\uL
        and a design that one can insert into any functioning
        cw ESR cavity.
    Most importantly, the method illustrated here allows us to intrinsically
        probe the sample heating and
        will permit the development of further improved systems.}
    \label{fig:t10andtemp}
\end{figure}
We measured the $T_{1,0}$ time (eq.~\ref{eq:t1fit})
    of water inside the new pass-through probe design
    as a function of incident microwave power.
The Hindman model (fig.~\ref{fig:hindmantemp})
    can then determine the sample temperature from the value of $T_{1,0}$
    (fig.~\ref{fig:t10andtemp}) at each microwave power increment.
These measurements (fig.~\ref{fig:Tvst}) confirm the
    significantly reduced sample heating with the pass-through
    probe design relative to the previously presented 
    open-bottom design from~\cite{Armstrong_portable}. 
Likely,
    both the more consistent positioning of the sample
    in the area of minimal
    electric field and 
    the more consistent air cooling
    contribute to the improved performance of
    the pass-through probe
    (cf. fig.~\ref{fig:t10andtemp}).
Furthermore (not shown in this plot),
    in repeated measurements,
    the pass-through probe presents a
    reproducible dependence of temperature
    on microwave power.
By contrast, upon removing and reinserting
    the NMR probe,
    measurements taken with
    the open-bottom design
    can vary significantly.
In fact, the variation of sample heating
    due to small changes in the sample positioning
    proves more problematic than the
    overall increase in the amplitude of sample heating itself.
This is because the
    irreproducible variations in heating
    seen with the open-bottom design
    block any attempts to
    systematically correct for
    the change in $T_{1,0}$ as a function of power,
    while one can correct for reproducible heating
    effects, as will be shown.

In summary, the $T_{1,0}$ of water
    allows us to identify the experimental setup
    that optimizes the $B_{1,\mu w}/E_{\mu w}$ ratio
    ($B_{1,\mu w}$ and $E_{\mu w}$ are the microwave
    magnetic and electric field amplitudes, respectively).
In consequence, this procedure allows one to optimize the
    ratio of the saturation, $s(p)$,
    to the dielectric heating.
However, even in an improved setup,
    the dielectric heating
    still introduces
    measurable changes in
    the bulk water relaxation time.
In turn, the theory predicts
    that these changes lead to
    measurable changes in the
    signal enhancements,
    especially for samples with
    low concentrations
    (\ie~$\le 500\uM$) of nitroxide spin probes.
\subsection{Observation of Enhancements and Relaxation Times}\label{sec:rel_times}
\outlineblank{subsubsection}
\begin{figure*}[tbp]
    \includegraphics[width=5in]{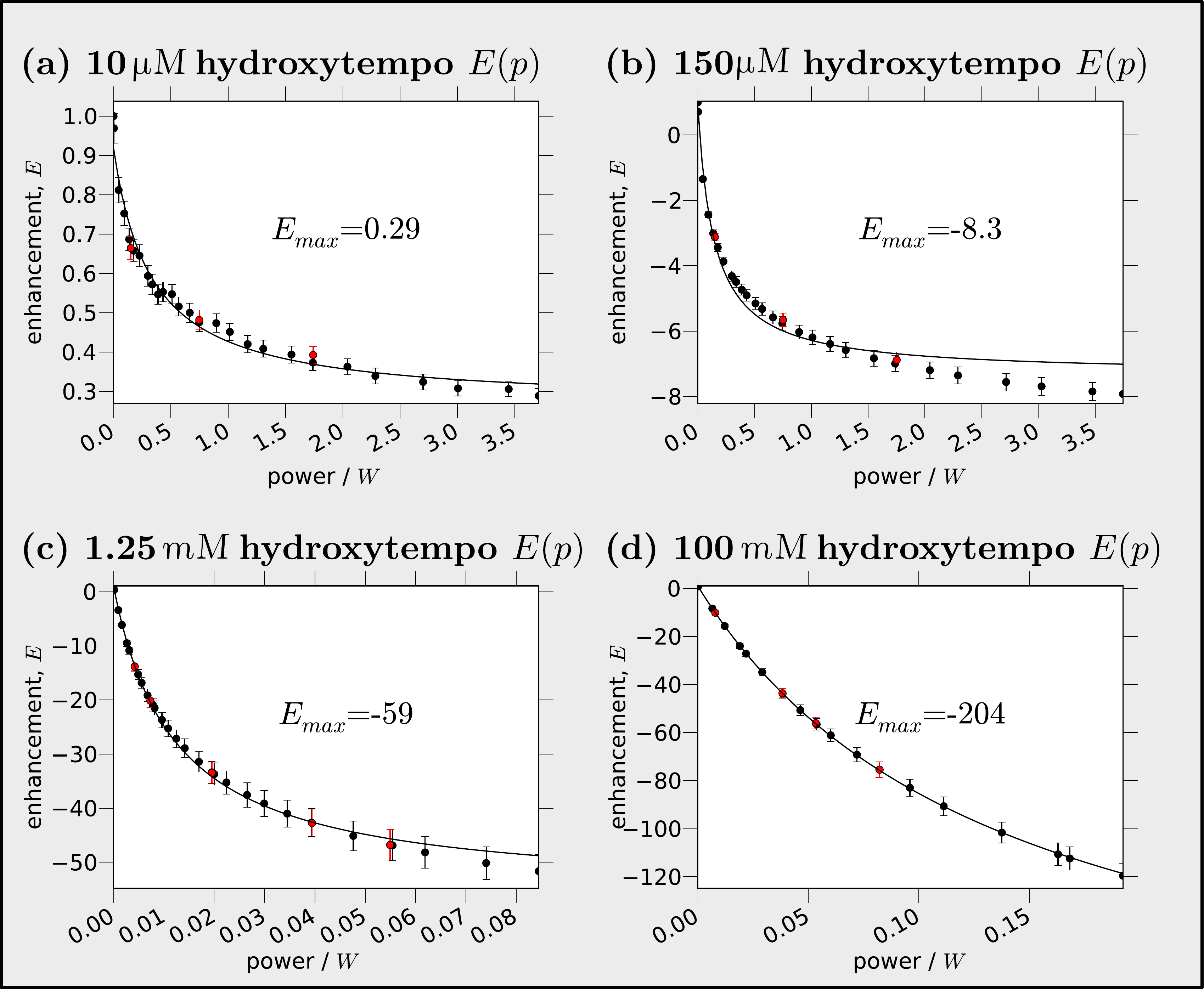}
    \caption{These plots show the ODNP signal enhancement, $E(p)$,
        as a function of power, $p$,
        for three different concentrations of
        \oht spin probes dissolved in water.
    We integrate the enhanced signal intensity
        and normalize it against the unenhanced signal intensity
        (\ie $E(0) = 1$)
        in order to determine the unitless value
        of $E$ displayed along the $y-$axis.
    The solid lines show the best fit to the uncorrected model
        (eq.~\ref{eq:polarization_subst}).
    The 100\mM sample follows the expected
        functional form,
        while the 150\uM sample clearly does not.
    Because of the residual misfit,
        we can only assign a rigorous meaning to
        the $E_{max}$
        for the 100\mM concentration
        sample, where it gives the value
        that the enhancements asymptotically approach
        with increasing microwave power.
    For all the data,
        the repetition delay, $t_{rd}$, between NMR scans satisfies
        $t_{rd}>5 T_{1,max}$ (see eq.~\ref{eq:t1max})
        in order to prevent
        artifactual suppression of the enhancements at higher powers
        (as in fig.~\ref{fig:Emax_mismatch}),
        while the dense sampling of data points
        plays a critical role in clearly identifying
        the misfit to the uncorrected model
        that occurs at the lower concentrations.
    The spectrometer acquires the black points in order of
        increasing microwave power;
        to check for reproducibility
        (e.g. lack of sample loss during the experiment),
        it then acquires
        the red (gray) points in order of decreasing
        microwave power.
        }
    \label{fig:representative_concs_emax}
\end{figure*}

Eq.~\ref{eq:f_is_linear} predicts that
    the enhancements observed
    for low concentration samples
    at higher microwave power
    will present a non-asymptotic (approximately linear)
    dependence on the microwave power.
As a result, they should deviate
    significantly from the uncorrected model (eq.~\ref{eq:polarization_subst}).

To test this prediction,
    we acquired the enhancement vs. microwave power,
    $E(p)$ (fig.~\ref{fig:representative_concs_emax}),
    curves for three different concentrations,
    150\uM, 1.5\mM, and 100\mM,
    of \oht spin probes freely dissolved
    in water\footnote{Here it is worth noting that
        the original cw ODNP measurements of the coupling factor~\cite{Armstrong_jcp}
        employed 4-oxo-TEMPO (i.e. tempone, 4-Oxo-2,2,6,6-tetramethyl-1-piperidinyloxy) solubilized in DMSO.
    As Bennati \etal more recently took advantage of~\cite{Turke_sat},
       4-oxo-TEMPO
       does have a very reasonable solubility in water.
    On the other hand, it also has a very high affinity for sticking
        to most glassware, making a reliable concentration series
        without DMSO problematic.}
    (fig.~\ref{fig:representative_concs_emax}).
At a 100\mM concentration of free spin probe,
    where $k_\rho C T_{1,0} \sim 100 \gg 1$,
    we expect the leakage factor to remain constant with power.
Thus, the uncorrected model (eq.~\ref{eq:polarization_subst})
    should still predict the enhancements accurately.
Indeed, the observed ODNP signal enhancements
    fit well to the expected asymptotic curve
    (depicted with the solid line),
    which extrapolates to an $E_{max}$ of -204.
This $E_{max}$ value agrees reasonably with recent literature data
    that draws from pulsed ESR and FCR measurements~\cite{Turke2010},
    while significantly exceeding previous predictions
    gleaned from cw ODNP measurements~\cite{Armstrong_jcp}.
However,
    for samples with 150\uM
    free spin probe concentration and lower,
    where $k_\rho C T_{1,0} \le 0.15 \ll 1$ (cf.~eq.~\ref{eq:f_is_linear}),
    even the best fit of the enhancements ($E(p)$)
    to the uncorrected model
    does deviate significantly
    (fig.~\ref{fig:representative_concs_emax}\myrefonefiftymicromolar),
    as predicted by eq.~\ref{eq:f_is_linear}.
\begin{figure*}[tbp!]
        \includegraphics[width=5in]{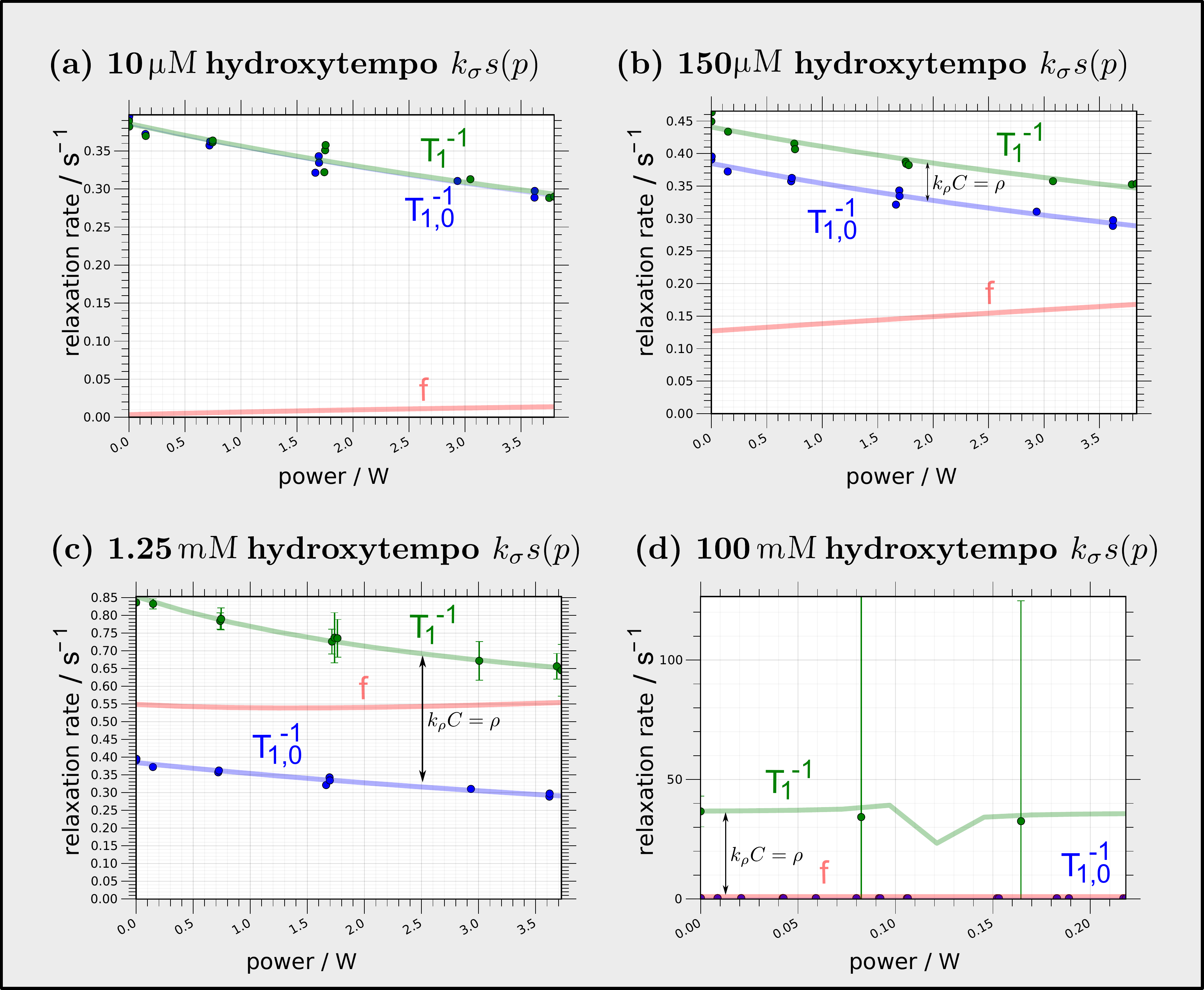}
    \caption{These plots show the variation of the bulk relaxation rate, $T_{1,0}^{-1}$, total relaxation rate, $T_1^{-1}$, and leakage factor, $f$,
    as a function of microwave power
    for four different concentrations of spin probe.
    We will shortly show that by accounting
        for the decrease in $T_1^{-1}$
        as a function of microwave power,
        the corrected model removes
        the misfit to the uncorrected model
        that we displayed in fig.~\ref{fig:representative_concs_emax}.
    These plots also allow one to track
        the dipolar self-relaxation rate,
        $\rho = k_\rho C$,
        which is
        the difference between
        the total ($T_1^{-1}$) and bulk ($T_{1,0}^{-1}$) relaxation rates
        and graphically appears as the distance between
        the two respective lines (black arrow).
    This data does not exclude the possibility
        that $\rho$ might slightly vary with power at
        high concentrations.
    However, at low concentrations,
        which are more important to biological studies,
        any change in the self-relaxivity,
        $k_\rho$, with increasing microwave power
        is insignificant compared to the change
        in $T_{1,0}$ and microwave power.
    By calculating the power variation of $f = \rho / T_1^{-1}$,
        we can see how dielectric heating
        impacts the leakage factor
        differently for different concentration of spin probe.
    At high spin probe concentration (100\mM),
        the leakage factor, $f$, does not vary significantly,
        since it has a value of $f \approx 1$ for all
        microwave powers,
        which is why the high probe concentrations demonstrate
        close adherence even to the uncorrected model (fig.~\ref{fig:representative_concs_emax}).
    By contrast,
        at low concentration,
        the leakage factor varies significantly,
        exhibiting an approximately linear dependence on
        power (as derived by eq.~\ref{eq:f_is_linear}).
        }
    \label{fig:leakage_diffconcs}
\end{figure*}

In a similar fashion, we can acquire the total
    relaxation rate, $T_1^{-1}(p)$, as a function of power,
    and compare it to the bulk water relaxation rate, $T_{1,0}^{-1}(p)$.
This allows us to estimate the leakage factor, $f$ (in fig.~\ref{fig:leakage_diffconcs}).
As predicted by eq.~\ref{eq:f_linear},
    $f$ changes as a function of microwave power
    for lower concentrations,
    while remaining consistently close to 1
    for all microwave powers at very high
    concentrations (approaching 100\mM,
        where $k_\rho C T_{1,0} \gg 1$).
The relaxation data of fig.~\ref{fig:leakage_diffconcs}
    also allows us to analyze the assumptions
    of the corrected analysis,
    so that we can understand these assumptions
    before proceeding to apply the corrected analysis
    to extract hydration dynamics information.
In particular, (like the uncorrected analysis)
    the corrected analysis
    does not account for
    how dielectric heating might alter
    the self-relaxivity, $k_\rho$,
    and cross-relaxivity, $k_\sigma$ (table~\ref{tab:symbol_table}).
We can measure the self-relaxation rate,
    $\rho = k_\rho C$,
    which (following eq.~\ref{eq:t1_vs_p_and_C})
    is the difference between
    the total relaxation rate, $T_1^{-1}$,
    and the 
    bulk water relaxation rate, $T_{1,0}^{-1}$,
    as shown in fig.~\ref{fig:leakage_diffconcs}. 
The observation of
    this value confirms an important feature
    of the corrected model.
Namely, at low spin probe concentrations ($< 2\mM$)
    the sizable variation in the bulk
    water relaxation rate ($T_{1,0}^{-1}$)
    with microwave power by far exceeds any variation of
    the self-relaxation rate
    ($\rho=k_\rho C$),
    so that any variation
    in $k_\rho$ with temperature
    is insignificant.

Surprisingly, these observations do
    contradict the intuitive notion
    that dielectric heating should primarily effect
    changes in the parameters traditionally regarded
    as containing all of the dynamic information
    relevant to the Overhauser effect:
    namely,
    the self-relaxivity
    (typically simply called the relaxivity), $k_\rho$,
    the cross-relaxivity, $k_\sigma$,
    and, the ratio between these two values,
    the coupling factor, $\xi$.
This is because
    uniformly, and especially
    at biologically relevant
    lower concentrations ($\le 500\uM$)
    of nitroxide probes,
    heating-induced variation in the relaxivities
    that encode the local dynamic information
    plays a far less important
    role in determining the ODNP signal enhancements
    than does heating-induced
    variation in
    the bulk water relaxation, $T_{1,0}^{-1}$.
Obviously, in cases where one employs cw ODNP
    to measure hydration dynamics
    with spin probe concentrations greater than 1\mM
    (where $k_\rho C$ begins to dominate
    the relaxation rates),
    or when more advanced instrumentation
    allows one can distinguish more subtle
    changes in relaxivity,
    one may wish to revisit variation of
    $k_\rho$ and $k_\sigma$ with microwave power
    as a secondary correction.
(Bennati \etal have already performed extensive modeling,
    based on FCR data~\cite{Bennati_fcr} that would assist
    in such an effort.)\footnote{However, we also caution that while
    the modes of sample motion that describe
    the motion of water near the nitroxide
    are likely equilibrated with the
    dielectrically excited modes of the bulk water,
    to the best of our knowledge,
    it has not yet been proven whether or not
    such an equilibration should take place
    under the experimental situation relevant
    to ODNP.
This is especially true since
    the sample resides in
    a small capillary tube,
    where the bulk water
    that interacts with the capillary is continuously cooled,
    while the timescales associated with the exchange and heat transfer
    between the bulk water and the hydration water are unknown.}
However,
    currently,
    the most important step in compensating for
    small, residual sample heating
    is that of correcting for the variation in the
    bulk water relaxation.
\subsection{Separate Calculation of $k_\sigma$ and $k_\rho$}
\begin{figure}[tbp]
    \begin{minipage}{3.5in}
        \begin{center}
            \includegraphics[width=3.5in]{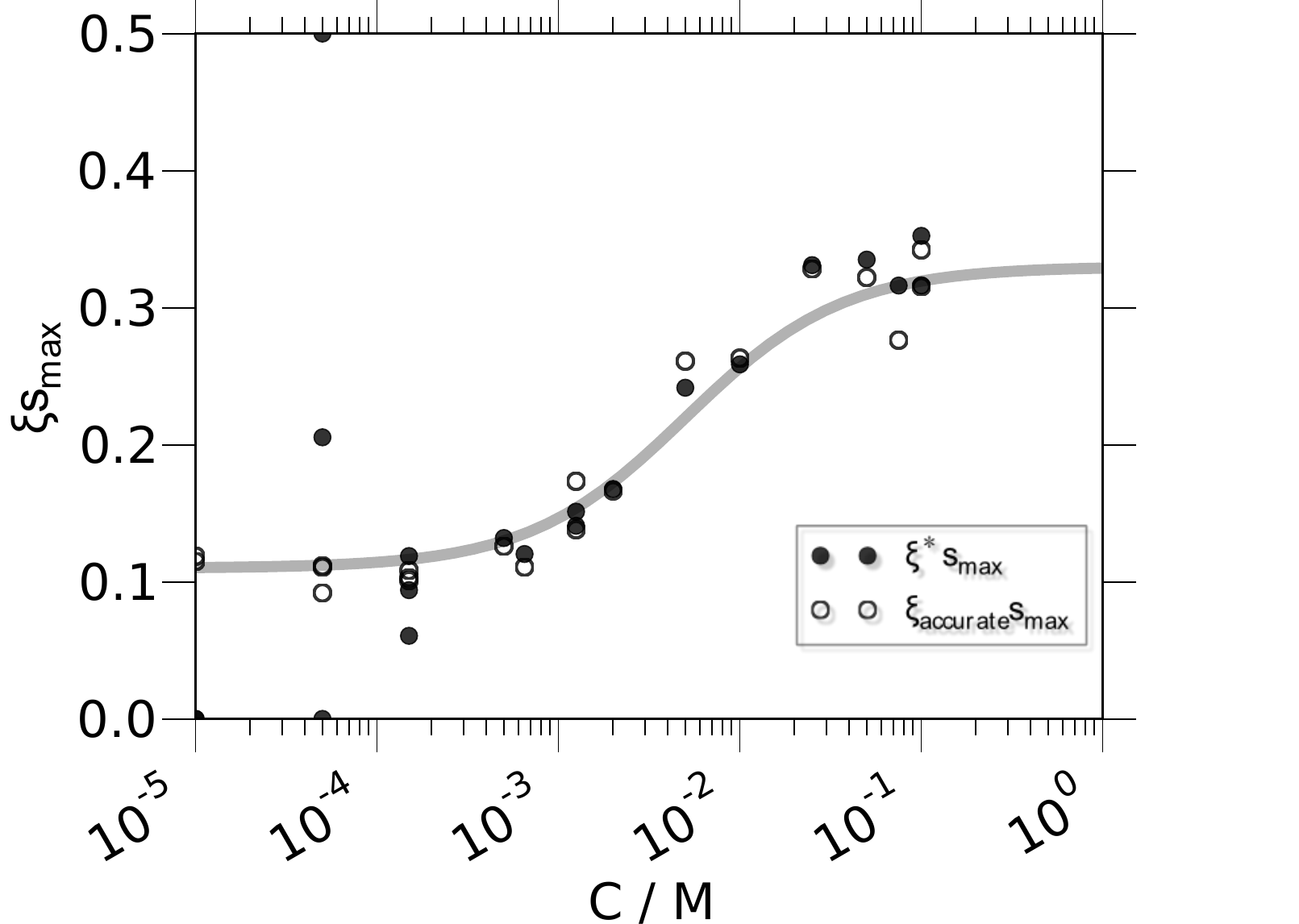}
        \end{center}
    \end{minipage}
    \caption{This figure presents a series of (unaveraged and uncorrected)
        measurements of $\xi s_{max} = \ksm/k_\rho$ calculated by two different methods.
    First, we determined $(1-E_{max})/T_1(0)$
        (\ie uncorrected $k_\sigma s_{max}$)
        and $k_\rho$ eq.~\ref{eq:k_rho_eq} from each sample
        independently.
    The ratio of the two yields the (closed circles) $\xi^* s_{max}$
        values presented, where we use the star to indicate
        that $\xi^*$ was extracted by the ill-determined method. 
    Note that, in the terminology of the standard
        uncorrected analysis,
        $\xi^* s_{max} = (1-E_{max})/f$.
    The resulting data has very high scatter at low concentration
        -- note in particular the outliers
        prevalent at lower spin probe concentration, $C$,
        of which the values lying at the limits of the $y$-axis
        have been thresholded, since the calculated values
        have such high errors that they are not physically realistic.
    In the second set of data ($\xi_{accurate} s_{max}$, open circles),
        we calculate $k_\sigma$ in the same way,
        but always use an average value of $k_\rho$
        determined from higher concentration data.
    The gray line gives \xiTurke~$s_{max}(C)$,
        where $s_{max}$ is given by eq.~\ref{eq:hyde_smax},
        with the value of $b''$
        as determine by Bennati \etal~\cite{Turke_sat}.
    Note that the value of $b''$ determines the concentration
        at which the inflection point (near a concentration of 4\mM)
        of the gray curve occurs;
        the fact that the data follows the curve through its inflection
        point means that the value of $b''$ is similar to that
        predicted by a ratio of the rates measured in~\cite{Turke_sat}.
        }
    \label{fig:k_rho_scatter}
\end{figure}

The study by Armstrong and Han~\cite{Armstrong_jcp}
    obtains $\xi$ by
    extrapolating a relatively evenly spaced
    series of concentrations between 0 and 15\mM,
    and is thus heavily weighted by the low concentration
    values.
This led us initially to 
    hypothesize that the difference between
    the higher coupling factor
    of \xiTurke~\cite{Turke2010}
    and the lower value of \xiArmstrong~\cite{Armstrong_jcp}
    arose from the fact
    that, at very low concentration,
    some finite population of bulk water
    does not diffusively exchange with the water
    near the spin probe on a timescale less than
    the NMR $T_1$ time;
this would result in having a continuum of different
    water populations with slightly different enhancement
    values, and would have the effect of lowering the
    enhancement at low concentration.
Rudimentary calculations led to the conclusion
    that this effect did not play an important role.

To investigate this issue further,
    we can ask what might be gained by separately calculating
    the cross-relaxivity, $k_\sigma$,
    and the self-relaxivity, $k_\rho$,
    as previously described in the theory section.
Fig.~\ref{fig:k_rho_scatter}
    shows the same raw data
    processed in two ways.
In one case,
    we calculate the product
    of the coupling factor
    and the saturation factor,
    which we denote with $\xi^{*} s_{max}(C)$
    (where 
    the star -- or lack of -- indicates
    the method of processing)
    to indicate an ill-determined $\xi^{*}$ value,
    which follows the calculations used in previous studies. 
For the second case,
    when we determine $\xi s_{max}(C)$
    via the corrected analysis,
    we acknowledge that $k_\rho$
    determined from the low-concentration samples
    has a very high percentage error;
the error comes from the fact that the
    total $T_1$ relaxation
    is dominated by the bulk water relaxation, $T_{1,0}$
    so that $k_\rho C$ is difficult to determine
    (cf.~eq.~\ref{eq:t1_vs_p_and_C} --
    because the difference
    between $T_1^{-1}$ and $T_{1,0}^{-1}$
    in eq.~\ref{eq:rho_at_power_off}
    is very small and is
    divided by a very small concentration, $C$).
Thus, a value for $k_\rho$,
    taken from an average of the higher concentration data,
    should be more accurate.
In fact,
    when we apply this value of $k_\rho$
    for the calculation of all $\xi s_{max}$ values at all
    concentrations, the scatter of the data is dramatically reduced,
    and even at concentrations as low as 10\uM,
    a clear and meaningful trend becomes apparent. 

Interestingly, without further correction,
    the shape of $\xi s_{max}$ vs. spin probe
    concentration matches the
    predictions of Bennati~\etal~\cite{Bennati_fcr,Turke_sat}.
Thus, this data unequivocally supports the
    higher value of \xireferenceval as the correct value
    for the coupling factor between water and freely dissolved
    spin label,
    as opposed to the lower value of \xiArmstrong
    previously used as a reference.
This data also appears to support a value of $b''/C$
    (\ie the ratio
    of the Heisenberg spin exchange and electron spin relaxation rates in
    eq.~\ref{eq:hyde_smax})
    that is at least the same order of magnitude
    as the value of $198.7\M^{-1}$
    determined from the data of T{\" u}rke \etal (\cite{Turke_sat}).

This leads to the remaining hypotheses that either the results
    of Armstrong and Han~\cite{Armstrong_jcp} are
    strongly affected by the presence
    of dimethyl-sulfoxide in their solution
    (\ie that it genuinely perturbs the hydration dynamics),
    that the artifact noted earlier
    (arising from the significant lengthening
    of the $T_{1,0}$ with microwave power)
    resulted in artificial suppression of the signals at high microwave powers,
    or that one or both of these effects
    combine with a
    large error when extrapolating
    measurements of $\le 15\mM$ to
    $s_{max}$ at high concentration.
Low concentrations
    of $\le 15\mM$ may not be a sufficiently high
    spin probe concentrations for extrapolating to
    high concentration,
    and may lead to large errors
    either when explicitly extrapolating
    $s_{max}$ to high concentrations,
    or when implicitly doing
    so by extrapolating \Emax to high concentrations,
    as done in~\cite{Armstrong_jcp}.
Regardless,
    these results support the conclusion
    that the higher measured value of
    the coupling factor, \xireferenceval~\cite{Bennati_fcr,Turke_sat,Turke2010},
    is indeed the correct value for the coupling factor
    between water and freely dissolved, small nitroxide spin probes,
    and that even rudimentary cw DNP instrumentation
    and analysis can quantify this value
    without the need for pulsed ESR and/or FCR instrumentation.
\subsection{Application of the Improved Analysis}\label{sec:impr_analysis}
\begin{figure*}[tbp]
        \begin{center}
            \includegraphics[width = 5in]{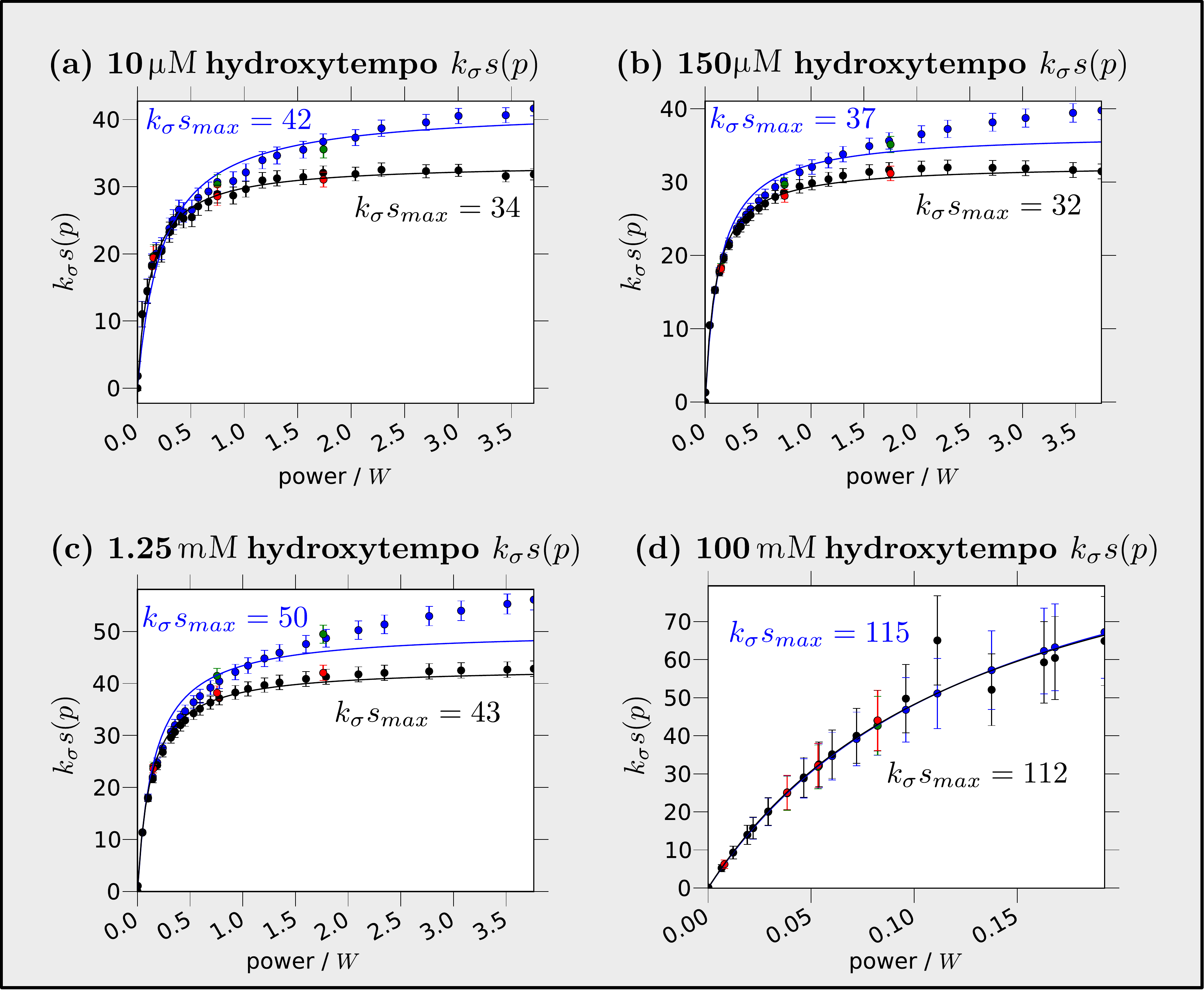}
        \end{center}
    \caption{
    These plots present the product of the cross-relaxation
        with the electron saturation factor, \ie \ksp,
        for the same series of experiments as fig.~\ref{fig:representative_concs_emax}.
    The uncorrected curves, in blue (gray),
        present $(1-E(p))/T_1(0)$
        and so have the same shape
        as the curves in fig.~\ref{fig:leakage_diffconcs}
        (though inverted);
    note that -- due to the imperfection in the uncorrected model --
        these are only {\it apparent} values of \ksp. 
    The black curves and points
        present the \ksp product,
        as determined by the corrected analysis
        (\ie $(1-E(p))/T_1(p)$ as determined
        by eq.~\ref{eq:polarization_subst_fundamental},\ref{eq:F_linear},\ref{eq:t1_from_F_linear}).
    With knowledge of the maximum possible saturation
        factor, $s_{max}$, one can extract the coupling
        factor directly from the fit shown
        (as in table~\ref{tab:xi_table}).
    The corrected data for the low concentration samples
        clearly level off at high microwave power,
    indicating that
        the highest microwave powers
        employed here
        significantly saturate the
        electron spin transition.
    Even at low concentration,
        the uncorrected data never level off,
        since heating induces a
        continuous increase in the leakage
        factor, resulting in a continuous increase
        in enhancements.
    The data for the higher concentration samples
        do not level off
        simply because ESR transitions
        require the application of
        more intense microwave radiation
        before they saturate appreciably
        (\ie \phalf is greater as a result
        of the fast Heisenberg exchange
        and subsequent broad ESR linewidth).
    The three sets of data fit to different values
        of the \ksm value because $s_{max}$
        remains closer to $1/3$ at low concentration,
        while at high concentrations,
        Heisenberg exchange drives $s_{max}$
        close to 1.
        }
    \label{fig:representative_concs_ksmax}
\end{figure*}

We can now analyze the difference between the 
    uncorrected analysis
    (eq.~\ref{eq:polarization_subst},\ref{eq:emax}),
    and the analysis that corrects
    for the dielectric heating effect
    (eq.~\ref{eq:polarization_subst_fundamental},\ref{eq:ksmax_extrap}).
As reviewed previously,
    whether one seeks to retrieve the
    translational correlation time, \correltime,
    or the local translational diffusivity of the hydration water, \Dlocal,
    one first needs to extract
    an accurate value for the coupling factor, $\xi$.
The uncorrected analysis
    employs the $E(p)$ curves and \Emax values
    of fig.~\ref{fig:representative_concs_emax}
    directly to calculate the hydration dynamics
    via eq.~\ref{eq:emax}.
However, in order to compare the two models 
    on an equivalent basis,
    we employ eq.~\ref{eq:polarization_subst_fundamental}.
For the uncorrected model,
    we plot $\ksp = \left( 1-E(p) \right)/T_1(0)$,
    where $E(p)$ is the data of fig.~\ref{fig:representative_concs_emax}
    and $T_1(0)$ is the NMR longitudinal relaxation time
    in the absence of microwave power.
We then measure $T_1(p)$, interpolated where necessary
    via eq.~\ref{eq:F_linear} and \ref{eq:t1_from_F_linear},
    as shown in fig.~\ref{fig:leakage_diffconcs}.
We insert the interpolated $T_1(p)$ values into
    eq.~\ref{eq:polarization_subst_fundamental}
    and plot the resulting corrected \ksp values
    in fig.~\ref{fig:representative_concs_ksmax}
    as well.

At 100\mM,
    and, to a lesser extent, at 1.25\mM spin probe concentrations
    the \ksp curves generated by both models
    match closely for all microwave powers
    (fig.~\ref{fig:representative_concs_ksmax}\myrefhundredmillimolar).
The high spin probe concentration
    leads to a fast
    self-relaxation rate,
    $k_\rho C$.
Since the self relaxation rate,
    $k_\rho C$,
    does not exhibit large changes with microwave power,
    it masks the change in the bulk water relaxation rate,
    $T_{1,0}^{-1}$,
    with microwave power.

At the lower concentrations of
    10-150\uM, the $T_1$ time changes
    significantly with microwave power
    (as seen in fig.~\ref{fig:leakage_diffconcs}).
Therefore,
    the apparent values of \ksp
    that the uncorrected model generates
    differ significantly from
    the corrected values of \ksp.
Most significantly, the data
    generated by the corrected analysis
    (eq.~\ref{eq:polarization_subst_fundamental},\ref{eq:ksmax_extrap})
    indeed level off
    visibly upon saturation of the
    ESR transition
    at high microwave power
    (fig.~\ref{fig:representative_concs_ksmax}\myreftenmicromolar-\myrefonefiftymicromolar),
    as predicted by the asymptotic
    model for ESR saturation
    (eq.~\ref{eq:smax}).
These corrected data should therefore
    remain consistent despite any changes in the hardware
    parameters, which would affect only $\Delta T_{1,0}$.
The corrected values of \ksp
    will also contribute less error associated
    with misfit to the model.
Finally, since the corrected values of \ksp
    level off at high power,
    we gain confidence that they approach the
    asymptotic \ksm value closely,
    and that extrapolation to infinite power
    will generate less error.
These two gains in accuracy become more significant
    at lower spin probe concentrations
    and at higher microwave powers.

We can also address the question of whether or not
    dielectric heating might alter the fundamental relaxation
    rate, $k_\sigma$.
Interestingly,
    the fact that all the data fit well to the corrected
    model implies that $k_\sigma$ does not vary
    significantly as a function of microwave power.

Table~\ref{tab:xi_table} collects the numerical
    results from these experiments.
Like in our initial analysis, these
    data support the higher coupling factor value
    of \xireferenceval.
We should note that additional experiments (data not shown)
    were done to verify that we choose a field relative
    to the cavity frequency that is positioned exactly
    on the ESR resonance.\footnote{\ie the field that maximizes \phalf}
We also note that the error analysis of the fitting
    procedure is relatively complex,\footnote{The typical
        covariance analysis of the asymptotic data becomes unusually
        complicated as a result of the fact that the correlation
        of the uncertainty between enhancement measurements
        is very high as a result of the normalization procedure.
        The development of a method to automatically process such data
        and compute such errors is currently underway.}
    and will be the subject of future publication.
\begin{table*}\renewcommand{\arraystretch}{1.1}
    \begin{minipage}{7in}
    {\scriptsize
    \newcommand\csty[1]{\makebox[1.5cm][c]{#1}}
    \newcommand{\tc}[1]{\multicolumn{2}{c}{---- #1 ----}}
    \newcommand{\gf}[1]{{\color{gray}(#1)}}
\begin{tabular*}{\linewidth}{@{\extracolsep{\fill}}rp{1.5cm}p{1.5cm}p{1.5cm}p{1.5cm}p{1.5cm}p{1.5cm}p{1.5cm}p{1.5cm}}
 \hline\hline
                                              & \multicolumn{2}{c}{10\uM} & \multicolumn{2}{c}{150\uM} & \multicolumn{2}{c}{1.25\mM} & \multicolumn{2}{c}{100\mM} \\
\cline{2-3}\cline{4-5}\cline{6-7}\cline{8-9}%
                                              & uncorrected               & corrected                  & uncorrected                 & corrected                     & %
 uncorrected                                  & corrected                 & uncorrected                & corrected\\
\cline{2-2} \cline{3-3} \cline{4-4} \cline{5-5} \cline{6-6} \cline{7-7} \cline{8-8} \cline{9-9}%
 \ensuremath{\ksm} / \relaxivityunits         & \csty{42}                 & \csty{34}                  & \csty{37}                   & \csty{32}                     & %
 \csty{50}                                  & \csty{43}                 & \csty{115}                 & \csty{112}\\
 \ensuremath{\smax}\cite{Turke_sat,HydeFreed} & \tc{0.334}                & \tc{0.352}                 & \tc{0.465}                  & \tc{0.968}\\
 \ensuremath{k_\sigma}                        & \csty{126}                & \csty{102}                 & \csty{105}                  & \csty{91}                     & %
 \csty{107}                                   & \csty{92.7}                & \csty{119}                 & \csty{116}\\
 \ensuremath{k_\rho} / \relaxivityunits       & \tc{\gf{-137} 353.4}      & \tc{\gf{391} 353.4}        & \tc{\gf{357} 353.4}         & \tc{353.4}\\
 \ensuremath{\xi}                             & \csty{0.36}              & \csty{0.29}               & \csty{0.30}                & \csty{0.26}                  & %
 \csty{0.31}                                 & \csty{0.26}              & \csty{0.34}               & \csty{0.33} \\
 \ensuremath{\xi^*} & \csty{\gf{-0.94}} & \csty{\gf{-0.74}} & \csty{\gf{0.27}} & \csty{\gf{0.23}} & \csty{\gf{0.30}} & \csty{\gf{0.26}} & \csty{\gf{0.34}} & \csty{\gf{0.33}}\\
 \taubulk / ps                                & \csty{26}                 & \csty{47}                  & \csty{44}                   & \csty{59}                     & %
 \csty{40}                                    & \csty{58}                 & \csty{31}                  & \csty{34}\\
 \hline\hline
\end{tabular*}
    }
    \end{minipage}
    \caption{
    This table gives the
        cross relaxivity, self relaxivity, and
        coupling factor
        for the dipolar interaction
        between water and free spin probes
        at ambient temperature,
        measured at four representative
        \oht spin probe concentrations.
    The gray numbers in parentheses
        present the inaccurate values
        that result from attempting to calculate
        $k_\rho$ from samples with very low
        concentrations of spin probe
        (\ie where eq.~\ref{eq:rho_at_power_off}
        has a significant error), as previously discussed.
    As also demonstrated in fig.~\ref{fig:k_rho_scatter},
        the resulting attempt to calculate a value of $\xi^*$
        gives data with sometimes physically unrealistic
        values (as seen here since $\xi$ must lie in the range
        $0<\xi<0.5$).
    Therefore, we focus on
        the data for the
        four representative
        \oht spin probe concentrations,
        where we have employed the value of $k_\rho$
        measured at 100\mM.
    We calculate a value of $\xi$ both
        with and without the correction
        for the change in the $T_1(p)$ times
        as a function of power
        (as given by eq.~\ref{eq:polarization_subst_fundamental}
        and fig.~\ref{fig:representative_concs_ksmax}),
        labeled as ``corrected'' and ``uncorrected,'' respectively.
    Note that at low concentrations,
        where the bulk water relaxation is more important,
        the correction is more significant.
    The various rows of this table outline the
        various steps of the corrected analysis,
        as given at the end of the
        theory section~\ref{sec:corrected_analysis_w_outline};
        the values of $s_{max}$
        are calculated via.~eq.~\ref{eq:hyde_smax},
        with $b''/C = 198.7\M^{-1}$.
    At all concentrations,
        the measured values of the coupling factor, $\xi$,
        more closely agree with
        the values of Bennati \etal~\cite{Turke_sat},
        as well as the values predicted by FCR~\cite{Armstrong_jacs,Bennati_fcr}
        and MD simulations~\cite{Sezer_xi},
        than those previously extrapolated
        by Armstrong \etal~\cite{Armstrong_jacs,Armstrong_jcp}.
    It is important to note
        that the difference in the coupling factor
        measured at different concentrations
        does not indicate whether or not
        the corrected data is accurate;
        rather correcting for the variation of $T_1(p)$
        with microwave power must always
        generate results that are more accurate than when
        such a correction is not performed.
        }
    \label{tab:xi_table}
\end{table*}

\begin{figure}[tb!]
    \begin{minipage}{\linewidth}
        \begin{center}
            \includegraphics[width=0.8\linewidth]{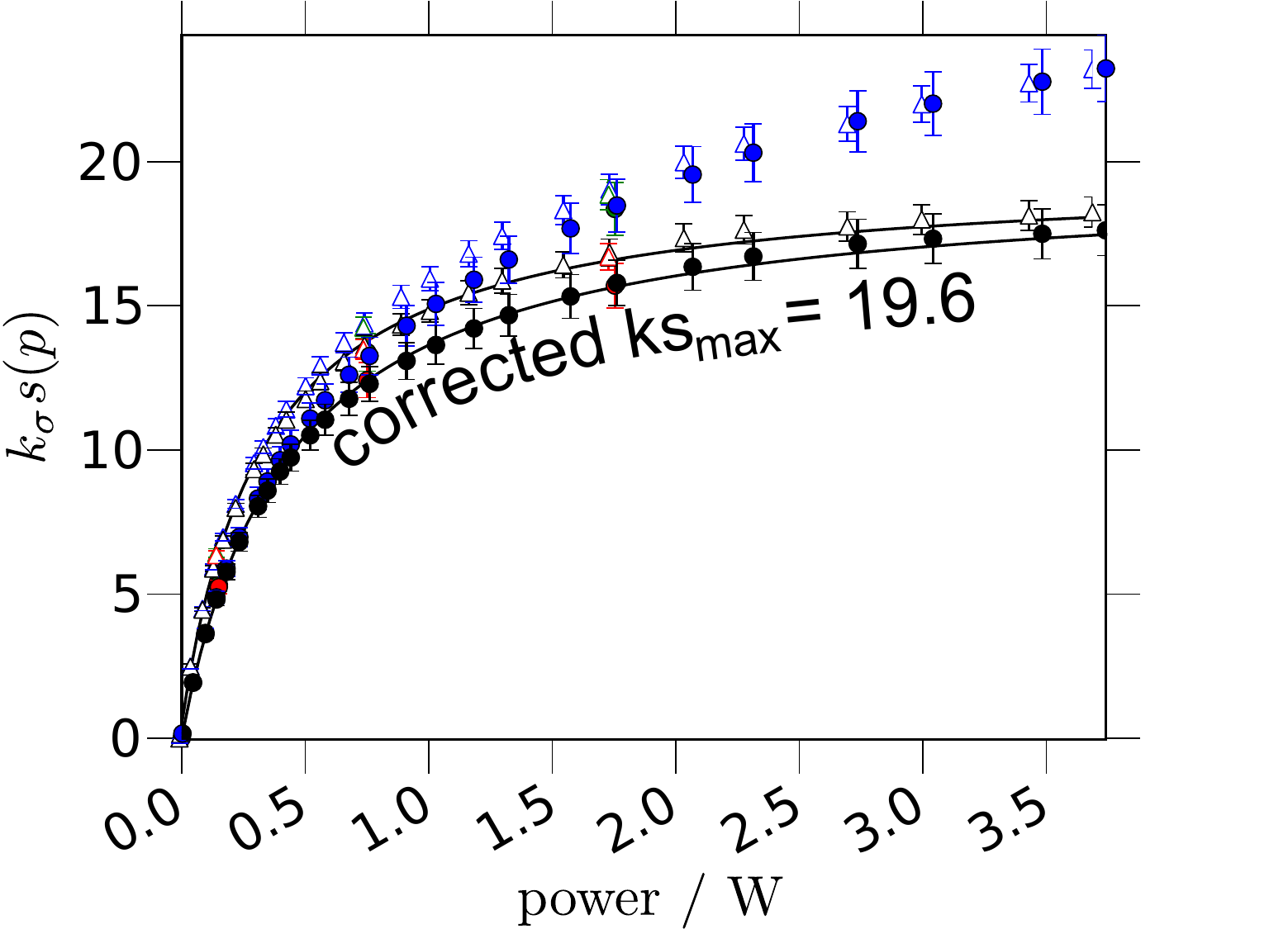}
        \end{center}
        \caption{This demonstrates the effect of the corrected analysis
            on an ODNP measurement taken for a mixture of
            tethered TEMPO label (3\% mole ratio) embedded in
            a DOPC lipid.
        Since a chemical bond attaches it to
            the very large vesicle membrane,
            we expect the spin probe to exhibit
            an $s$ of near 1 (as shown in \cite{Armstrong_jcp})
            leading to $k_{\sigma} s_{max}\approx k_{\sigma}$.
        This is confirmed by the fact that
            a solution containing 32.0\mM lipids (circles) and
            a 4.89\mM lipids (open triangles),
            \ie 9.6\mM and 150\uM of spin probe,
            yield closely agreeing $k_\sigma s_{max}$
            values of 19.6 and 19.5, respectively.
        The values plotted here are scaled by the $T_1$,
            following eq.~\ref{eq:polarization_subst_fundamental}.
        Therefore, despite the similar appearance of the two lines,
  the 32\mM sample does indeed exhibit much
            higher values of enhancements
            ($\Emax\approx -20$)
            than the diluted sample
            ($\Emax\approx -4.9$).
        For clarity, we have omitted the best fit lines
            for the uncorrected data in blue (gray),
            which -- like for the free label --
            exhibits misfit beyond the bounds of the experimental error.
        The best fit values for the uncorrected \ksm are
            24.5\relaxivityunits for the 32\mM DOPC sample
            and 23.2\relaxivityunits for the 9.6\mM DOPC sample
            (we use the average value of 23.9\relaxivityunits
            in the analysis in the text).
        The slight difference in the powers at which the two
            samples saturate likely indicates a slight difference
            in the $B_1$ conversion ratio (related to the cavity $Q$ factor)
            after loading the two samples.
        }
        \label{fig:bioexample}
    \end{minipage}
\end{figure}
\subsection{Application to a Sample with Tethered Spin Probes}\label{sec:application_of_anal}
An application of the new analysis to an
    example biological system helps to clarify
    the importance of the preceding results.
We selected 200\nm diameter unilamellar vesicles made of DOPC
    lipid\footnote{18:1 ($\Delta$9-Cis) PC (DOPC),
        \ie  1,2-dioleoyl-sn-glycero-3-phosphocholine,
        Avanti Polar Lipids \#850375}, dissolved in water at
    two different concentrations:
    4.89\mM and 32.0\mM.
We perform inversion recovery measurements
    to determine the value of $T_{1,0}(p)$ as a function of
    microwave power, $p$.
For the ODNP (\ie $E(p)$) and $T_1(p)$ measurements,
    we prepare another sample that also includes
    3 mole \% of a lipid with a tethered
    nitroxide probe located at the surface of the lipid bilayer.\footnote{16:0 Tempo PC,
        \ie~1,2-dipalmitoyl-sn-glycero-3-phospho(tempo)choline,
        Avanti Polar Lipids \#810606}

Biologically relevant systems
    with tethered spin probes
    tend to give results that one can interpret 
    in a more straightforward manner than
    one can interpret the results
    from freely dissolved spin probes.
Specifically,
    because the tethering to larger biological systems
    restricts the TEMPO spin probe dynamics,
    we can assume that nitrogen relaxation
    drives $s_{max}$ to closely approach 1,
    as explained in~\cite{Armstrong_jcp}.
It follows that the extrapolated value of \ksm
    conveniently approximates
    the cross relaxivity, $k_{\sigma}$.

At the same time,
    the same motional restriction also typically
    causes tethered spin probes to
    exhibit a faster ESR relaxation
    time than untethered spin probes
    and thus makes the ESR transition harder
    to saturate.
Specifically, notice how \ksp for
    a free spin probe
    of 150\uM concentration
    in fig.~\ref{fig:representative_concs_ksmax}\myrefonefiftymicromolar
    begins to level off near 1~W of microwave output power,
    while the \ksp values for the
    4.89\mM sample of DOPC
    with a spin probe concentration of 150\uM
    do not begin to level off until about 3~W,
    as presented in fig.~\ref{fig:bioexample}. 
The experiments on DOPC here
    employ approximately the same range of power
    as the
    free spin label studies presented earlier
    in this publication.
Arguably, the free spin label studies
    exceed the powers strictly necessary
    to achieve clear ESR saturation
    at low concentrations
    when they employed microwave output powers
    of up to 2-3~W.
However, for the ODNP study of the lipid vesicle surface,
    enhancements must be acquired
    at output powers approaching 3~W
    before saturation becomes evident.
The use of such high powers
    can lead to rather significant heating
    effects,
    making the corrected analysis even more important
    to quantitatively determining the coupling factor.
In the case of the DOPC sample,
    we observe an
    uncorrected value of $k_\sigma s_{max} \approx k_\sigma$
    that differs from the
    corrected value by more than 20\%.

Like in the free spin probe study,
    in order to avoid using an ill-determined leakage factor,
    we do not calculate $\xi$
    for the uncorrected analysis
    from $\left( 1-\Emax \right)/\polratio f$.
Rather, we first determine the
    value of $k_\sigma$ for both the corrected
    and uncorrected analyses.
Then, we compare the $T_1$ times of the
    labeled and unlabeled samples
    in the absence of microwave power
    from the 32\mM sample,
    namely 314\relaxivityunits,
    in order to retrieve
    a reliable value of $k_\rho$ that we can use 
    to calculate the coupling
    factor from eq.~\ref{eq:coupling_factor} ($\xi = k_\sigma/k_\rho$),
    for both the high and low concentration
    samples (following eq.~\ref{eq:k_rho_eq}).
We retrieve
    a corrected value of the coupling factor $\xi = 0.062$, while we retrieve an uncorrected
        value of 0.076. 

Similarly, we can determine the value of \tausite corresponding
    to \xsite, which is measured at
    a local site in or on the macromolecule
    where the spin probe is tethered.
The corrected
    vs. uncorrected analyses
    measure different values for \xsite
    (fig.~\ref{fig:bioexample}),
    leading to correlation times of
    226\ps vs. 263\ps, respectively.
In previous work, researchers
    determined \xsite based on the uncorrected analysis;
    through the discussions and demonstrations presented here,
    it is clear that the
    corrected analysis should be used instead.
The correction of the value of \xsite implies a measurable
    correction for the local
    translational diffusion coefficient. 
Insertion of the two values for \tausite
    into eq.~\ref{eq:calcD}
    determines that, in this case,
    the two values of $D_{local}$ generated
    from the corrected and uncorrected
    analyses differ by more than 16\%.
Since other samples with tethered spin probes
    should present similar ESR and $^{14}$N relaxation times,
    we expect them to exhibit similar differences.
Thus, previous ODNP studies presenting relative trends for the
    \xsite, \tausite or \Dlocal values remain valid,
    but their absolute values need to be cautiously reexamined,
    as will be further detailed.

One should remain cautious
    that even the corrected method laid out here
    might not apply in a straightforward fashion
    to chemical systems where
    the small residual changes in temperature
    with increasing microwave power
    might induce a transition
    in the structure and dynamics of the sample.
Fortunately, most or all previously investigated samples
    that one would suspect of presenting such problems --
    such as
    macromolecules that undergo a glass transition~\cite{Kausik_pccp,Cheng_Han_Lee_jmr},
    or proteins that undergo aggregation~\cite{Pavlova_pccp}, folding, or unfolding~\cite{Armstrong_apomb}
    --
    the transition temperature does not fall
    between room temperature
    and the temperature at the maximum microwave power ($\approx 35^oC$). 
However, even in other systems that 
    undergo transitions within the experimental temperature range,
    the methodology presented here
    should remain very useful by
    registering and highlighting any sudden changes in $k_\rho$ and/or $\xi$
    as an unexpected change
    in \xsp (\ie misfit to the corrected model)
    with increasing microwave power, $p$.
\subsection{The Corrected Analysis in Context}\label{sec:impr_incontext}
\begin{figure}[tb!]
    \begin{center}
        \includegraphics[height = 3in]{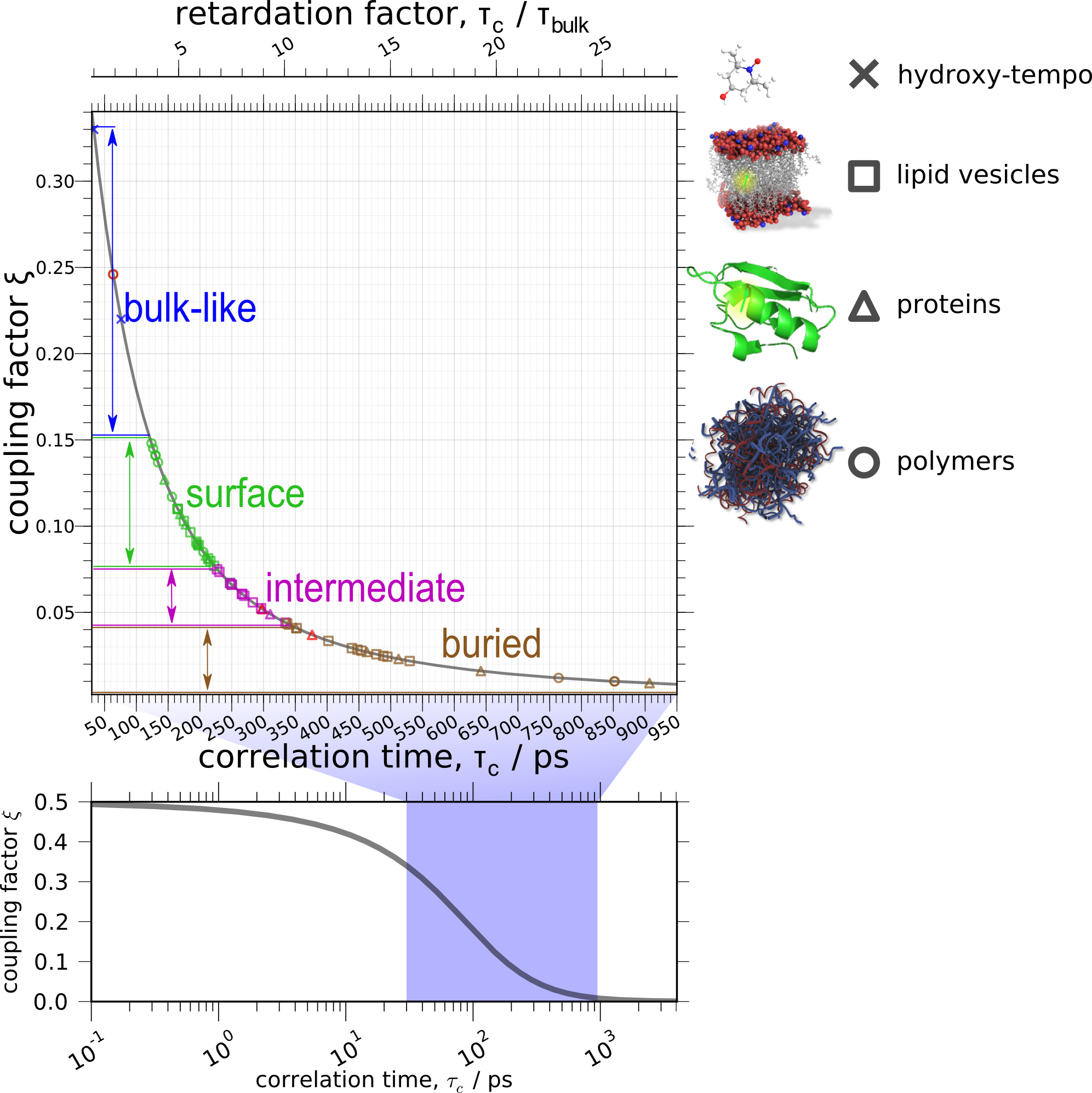}
    \end{center}
    \caption{Here, we collect data for the coupling factor, $\xi$,
        between a spin probe attached to specific, known biological
        sites and the nearby (within 5-10\Ang) hydration water.
    Even though these data come from previous publications that
        employ the uncorrected analysis,
        the results reproducibly
        fall within a particular zone corresponding to the
        spin probe location (\ie~exposed, intermediate, or buried
            as defined in the text).
    The exceptions to this trend are the lipid.
The vesicle doxyl-stearic acid labels (not shown) consistently show
        significantly higher dynamics
        than their tempo-PC analogs (which are presented in this plot).
    Hyaluronic acid~\cite{Ortony_njp},
        a polyelectrolyte
        which exhibits exceptionally fast, \ie bulk-like, dynamics.
    The unfolded protein mfp151~\cite{Ortony_njp},
        exhibits buried-like dynamics,
        and because of this is believed to have a some amount of local structure near the spin labeled site~\cite{Ortony_njp}.
    The native (\ie folded) state V66R1 site in ApoMb (Apo-myoglobin)~\cite{Armstrong_apomb} exhibits intermediate dynamics,
        even though it is believed to be on the surface of the protein.
    The retardation factor $\tau_c/\taubulk = \Dw/\Dlocal$
        gives the slow-down of the translational hydration dynamics relative
        to bulk water;
    the measurement for \taubulk sets the value of 1 on the
        retardation axis and, therefore, for the absolute values of the diffusivities.
    Both the correct value (at $\tau_c/\taubulk = 1$)
        and the previously employed value
        (at $\tau_c/\taubulk = 2.3$)
        are shown above with a blue ``x'' above.
    Of course, if the previously employed value were used to set the value
        of 1 on the retardation axis, the value of the diffusion
        would change for all the data points shows.
        }
    \label{fig:zone_figure}
\end{figure}

In this section, we can now analyze how the new methodology
    will lead to a reinterpretation or adjustment of previously
    acquired results.
Specifically, we will present how corrections
    for sample heating can lead to changes in the values
    determined for the coupling factor, $\xi$,
    in a given study.
Here, an important parameter of biophysical significant is the
    retardation factor of the ODNP-measured translational
    diffusion dynamics near a spin probe functionalized onto
    a biomolecular system, relative to that of bulk water.
In order to extract such retardation factors,
    we need to accurately determine the coupling factor
    between the freely dissolved spin probes and bulk water,
    which we use to quantify the translational correlation time,
    \taubulk.
Past studies have had much greater difficulty
    in employing the uncorrected analysis
    to extract consistent values for
    the translational correlation time of water
    interacting with freely dissolved
    spin probes than in employing the uncorrected
    analysis to extract consistent values for
    the translational correlation time of water interacting with
    spin probes tethered to macromolecules.
Previous ODNP dynamics studies (excluding~\cite{Ortony_njp,Cheng_Han_Lee_jmr})
    employ a relatively low value of \xiArmstrong
    as the reference value for the coupling factor of bulk water.
Eq.~\ref{eq:xi_fn_of_J} implies that
    this corresponds to a translational correlation time
    of $\taubulk=\tauArmstrong$.
However, the results here clearly support the higher
    bulk water coupling factor
    of \xireferenceval,
    which rather implies (via.~eq.~\ref{eq:xi_fn_of_J}) a choice
    of \taubulkequation.
Eq.~\ref{eq:calcD} points out that
    the local diffusivities, \Dlocal,
    for all soft matter and biological samples
    are approximately proportional to the ratio of
    the coupling factor in freely diffusing water, \taubulk,
    to the coupling factor within
    a local volume around the tethered spin probe
    in a particular sample, \tausite.
Therefore, a choice of the higher (and correct)
    reference value of the bulk water coupling factor
    means that the values of \Dlocal
    for all samples are about two-fold
    slower relative to bulk water than was previously thought.

Given the existing controversy on the bulk water coupling factor,
    most ODNP studies in the literature
    anticipated such a dilemma.
Therefore, they reported local hydration dynamics in terms
    of the translational correlation times, \tausite,
    rather than in terms of the local hydration water diffusion
    coefficient, \Dlocal,
    since the value of \tausite does not
    depend on the bulk water coupling factor.
Since we have found unequivocal agreement
    that the bulk water coupling factor
    is $\xbulk = \xireferenceval$,
    we are now at liberty to use either
    \tausite or \Dlocal,
    to describe the site-specific local hydration dynamics,
    depending on the context.
However, as eluded to before,
    comparisons between
    the value of \Dlocal
    for the local translational hydration dynamics
    determined by different
    techniques are valid,
    while such comparisons based on the absolute
    value of \tausite are not necessarily valid.

The lower concentration ($\le 1.25\mM$) values in 
    table~\ref{tab:xi_table} exhibit differences
    between the corrected and uncorrected values
    of 15 - 20\%,
    while, as noted before,
    the DOPC sample exhibits a heating
    correction of greater than 20\%.
Therefore, we expect if samples
    were re-analyzed with the corrected analysis,
    the value of the measured coupling factor would
    decrease by up to 20\%, though this number may vary.
On the one hand, the pass-through probe
    employed here significantly minimizes
    the potential amount of net heating in the system. 
On the other hand,
    the microwave powers
    and the Q of the cavity
    employed in previous studies
    likely varies with each study,
    and are not necessarily as high as those used here.
Thus, one can predict
    that moving an experiment to a different instrument,
    or even positioning the NMR probe slightly differently,
    can easily induce changes in $\Delta T_{1,0}$.
Either changes in $\Delta T_{1,0}$, or changes
    in the microwave powers sampled
    could cause the measured values of \xsite
    to vary.
Furthermore, each different protocol
    may or may not have employed
    sufficiently long recycle delays
    (\ie delays that exceeded $5\times T_{1,max}$ of eq.~\ref{eq:t1max}).
Therefore, in addition to a systematic decrease
    in the measured values of \xsite,
    the corrected analysis may lead to a decreased scatter in
    the coupling factor value for a given sample,
    especially for samples with low spin concentrations.
Because the heating correction
    is on the order of 20\% for the well-controlled 
    hardware setup shown here, and -- as previously noted --
    we expect the previous setups to be significantly
    less well-controlled,
    we believe that a conservative estimate
    of the scatter in the measurement of the coupling
    factor, \xsite,
    is about 5\%.

The corrected analysis should
    remove any dependence on $\Delta T_{1,0}$
    to allow one to determine
    an \textit{absolute} value of \tausite and \Dlocal,
    and to accurately and reproducibly
    reduce the scatter across different protocols and
    instruments.
However, despite the previous inaccuracy
    in the absolute values
    of \Dlocal, previous publications
    reproducibly employ measurements
    of the coupling factor
    to identify meaningful and biologically significant
    trends and transitions
    associated with changes in the 
    local hydration dynamics~\cite{Ortony_njp,Cheng_Han_Lee_jmr,Armstrong_apomb,Pavlova_pccp,Kausik_pccp,Kausik_jacs}.
Since these studies employ the uncorrected analysis,
    the successful identifications that they
    achieve depend on the consistency
    of various experimental parameters.
In particular, the researchers who
    acquired the datasets for
    previous measurements
    empirically discovered the need to
    only compare coupling factors determined
    in precisely the same position in the same
    cavity, from enhancements
    acquired over the same sampling of
    microwave powers.
Such an experimental scheme
    ensures relatively consistent heating effects,
    \ie a relatively uniform dependence on $\Delta T_{1,0}$.
Therefore,
    it is crucial to emphasize that
    all qualitative trends and comparisons
    \textit{within each dataset},
    as previously published,
    should remain after correction. 

Finally, to alleviate concerns over
    comparing data that come from different
    studies (\ie datasets),
we analyze a subset of
    previously published studies
    to estimate an upper limit on the error
    associated with comparing these different studies.
It is worth noting that most of the data
    we review here
    are acquired on the same microwave cavity and NMR probe
    design and come from samples with relatively
    high, $\ge 500\uM$, spin probe concentrations,
    where the correction to the \xsite values 
    (and, therefore, the \tausite and \Dlocal values)
    should be smaller than for lower concentration
    samples.
For all these samples
    it is also known whether the
    spin probe resides
    within the hydrophobic core of a macromolecule
    or near the surface of a macromolecule.

When we compare this collection of measurements (fig.~\ref{fig:zone_figure}),
    we find that we can sort the dynamics into four categories.
With the exception of the polyelectrolyte hyaluronic acid,
    which exhibits very fast
    hydration dynamics~\cite{Ortony_njp},
    bulk water is the only sample to exhibit
    $\xi>0.15$.
Thus, we can classify this zone ($\xi > 0.15$) as ``bulk-like,''
    \ie significantly faster than the dynamics
    observed on any labeled compounds.
Researchers have routinely observed
    values of the coupling factor $0.075<\xi<0.15$
    for surface dynamics of macromolecular systems, including
    unfolded proteins and
    uncomplexed polyelectrolytes,
    as well as the surface of lipid vesicles,
    and folded globular proteins~\cite{Ortony_njp,Armstrong_apomb,Cheng_Han_Lee_jmr,Kausik2009b,Kausik_jacs,Kausik_pccp}. 
Researchers have routinely observed
    value of the coupling factor $0.042<\xi<0.075$
    for samples where one expects intermediate dynamics,
    including surface labeled vesicles
    in the presence of viscogens
    such as 20\% PEG~\cite{Ortony_njp} or binding agents
    such as $\ge 35\uM$ P188~\cite{Cheng_Han_Lee_jmr},
    lipid vesicles with spin labels
    attached at the carbon positions
    immersed five to ten C-C bonds into the lipid bilayer
    (\ie for label molecules 10 Doxyl PC,
    7 Doxyl PC, and 5 Doxyl PC)~\cite{Kausik_pccp},
    and the surface of the partially folded
    molten globule state of ApoMb~\cite{Armstrong_apomb}.
Researchers have routinely observed
    value of the coupling factor $\xi<0.042$
    for samples where one expects
    the spin labeled site to be buried inside
    a macromolecular complex;
such samples include complexed polyelectrolytes~\cite{Kausik2009b,Ortony_njp},
    aggregated protein fibrils~\cite{Ortony_njp,Pavlova_pccp},
    buried sites of natively folded
    and molten globule states of
    ApoMb~\cite{Armstrong_apomb},
    and lipid vesicles
    where the spin label is attached
    14 bonds deep into the vesicle
    (\ie 14 Doxyl PC)~\cite{Kausik_pccp,Cheng_Han_Lee_jmr}.

Interestingly,
    at this level of resolution,
    the location (\ie~surface vs.~buried) of the site under investigation
    seems to be the primary determinant of
    the translational hydration dynamics,
    regardless of whether the site resides on a protein,
    membrane vesicle, or polymer.
Thus, the cw ODNP method can already 
    \textit{classify the location of the site}
    based on its hydration dynamics (fig.~\ref{fig:zone_figure}).
With the improvements presented here,
    future studies will reproducibly resolve
    and classify sub-categories of dynamics, 
    allowing us to map out the hydration-dynamics-based
    landscape of proteins,
    lipid vesicles, and other soft matter systems
    on a finer, absolute scale.
Previous research suggests that
    changes in the surface hydration dynamics
    should correlate strongly with the interfacial forces
    that drive biological and polymeric transitions~\cite{Frauenfelder2009,heyden2010dissecting,Zhong_cpl,CremerReview,patel2010fluctuations}.
As seen from the ApoMb and vesicle data,
    interesting effects,
    such as protein folding and complexation,
    initiate at sites with intermediate-regime
    hydration dynamics.
Therefore, there is particular interest
    in such high resolution investigations
    into the rich variation of the hydration dynamics
    within the intermediate regime.
\section{Conclusions}
\outlineblank{subsection}
\outlineblank{subsubsection}

The corrected analysis proposed here
    (eq.~\ref{eq:polarization_subst_fundamental},\ref{eq:ksmax_extrap}) represents a significant improvement
    in the ability of cw ODNP to accurately quantify hydration dynamics.
It is the first model to
    fit accurately acquired
    (\ie without artifacts)
    enhancement vs.~power, $E(p)$, data
    for low ($\le 500\uM$) concentrations of spin probes.
The data also resolves the debate over the determination
    of the bulk water coupling factor,
    pointing out that the higher value
    (\xiTurke, \xifcrBennati, \xifcrArmstrong, and \ximd from
    \cite{Turke_sat}, \cite{Bennati_fcr}, \cite{Armstrong_jacs}, and \cite{Sezer_xi}, respectively),
    rather than one previously
    determined by Armstrong \etal
    (\xiArmstrong from~\cite{Armstrong_jacs}),
    should be used as the reference value for
    measurements of hydration dynamics.
Entirely cw ODNP instrumentation
    can retrieve the resulting high \Emax values,
    without the need for pulsed ESR instrumentation,
    despite previous predictions~\cite{Armstrong_jacs,Armstrong_jcp}.
From one point of view, this agreement supports the validity of the FCR, MD,
    and pulsed ESR measurements, which employ similarly high spin
    probe concentrations. 
From another point of view, it supports the validity of
    the experimental strategy
    proposed in~\cite{Armstrong_jacs},
    which employs solely cw ODNP hardware
    as a viable quantitative analytical
    tool for determining the coupling factor, $\xi$.
This conclusion
    is particularly promising and advantageous
    to the ODNP and analytical biochemistry communities
    since, as previously mentioned,
    cw ODNP hardware requires
    significantly less user training
    and expenditure than pulsed ESR hardware,
    and thus is more broadly available.

The results presented here outline standard tests
    that will facilitate further hardware development
    and clarify how one can proceed further
    in the task of developing cw ODNP as 
    a generally applicable, push-button analytical
    tool for the quantification of local hydration dynamics.
For the development of new NMR probes and/or ESR cavities,
    the measurement of the longitudinal NMR relaxation
    of bulk water, $T_{1,0}(p)$,
    as a function of microwave power, $p$,
    gives a complete test of the hardware's ability
    to regulate the sample temperature in the presence
    of the strong microwave fields.
However, the results actually support
    a better and faster standardized test of
    newly implemented hardware systems
    -- namely,
    the corrected cw ODNP analysis for $\le 150\uM$ \oht.
The enhancements and relaxation times
    of such low concentration systems
    change significantly in order to clearly indicate
    any heating effects (as in fig.~\ref{fig:representative_concs_emax}\myreftenmicromolar),
    while at the same time indicating the
    relative power conversion ratio (inversely proportional
    to $\sqrt{p_{1/2}}$ on resonance)
    of the system.
Therefore, this standardized test
    will give a clear and quick indication of
    both how much power
    the system requires before
    it will appreciably saturate the ESR transition,
    as well as
    how much dielectric heating
    the hardware subjects the sample to
    at the required powers.

Five clear conclusions dominate this presentation:
    by measuring the bulk relaxivity,
    one can intrinsically probe the sample temperature
    and thus develop the next generation
    of cw ODNP hardware;
    by measuring and accounting for $T_{1,max}$,
    one can acquire reproducible signal enhancement data for
    a particular system;
    by implementing the new analysis presented here,
    one can recover accurate values of $\xi$, even
    in the presence of moderate sample heating;
    by repeating measurements at high concentration
    carefully,
    cw ODNP can retrieve values of the coupling factor
    that do indeed agree with predictions given by pulsed ESR
    and FCR measurements taken at similar concentrations;
    finally,
    by independently measuring $k_\sigma$, 
    one can access information about the translational
    dynamics, even at concentrations of less than a hundred
    micromolar, opening up opportunities to study
    a wide range of new systems of biological significance.
In addition to the method of calculating the self-relaxivity,
    $k_\rho$, at high concentration in order to determine the
    coupling factor, $\xi$,
    this also opens up the possibility of independently verifying
    the concentration
    (for instance, by UV-visible or IR spectroscopy)
    and observing trends in a fashion similar to how trends
    are currently observed in the coupling factor.

The results here also make it clear what comes
    next in the task of developing cw ODNP as an
    analytical tool for the quantification of local
    hydration dynamics.
Systematic comparisons between
    measurements taken on different instruments
    or microwave cavities should now
    agree quantitatively.
In particular, experiments that observe the change in the
    coupling factor across the various frequencies possible
    with X-band equipment
    would offer interesting insight
    into how accurately the existing models
    describe the hydration dynamics
    and should be possible with tunable cavity setups
    (similar extensions for the S- through Q-bands
    would be similarly interesting).
However, until now,
    an understanding of ODNP that was capable of
    reproducibly extracting dynamics
    under variable conditions of sample heating and
    cooling was lacking,
    while such variations inevitably occur in a
    cavity with a tuning range of several \GHz
    (compare to~\cite{Armstrong_portable}).
This study also paves the way for further improved
    methodologies that compensate
    for sample heating, if they prove necessary.
As previously discussed, even the significant improvements
    developed here might not suffice for some
    particularly difficult systems.
For such systems,
    a properly integrated and automated cryostat
    can adjust air surrounding
    the sample to a lower temperature that compensates
    for the microwave heating.
One can iteratively optimize
    such a compensation against the criterion that the
    $T_{1,0}$ remains constant with increasing
    microwave power.\footnote{$T_1$   measurements should be sufficient for samples with low concentration of spin probe, where $T_{1,0}^{-1}$ is the same order of magnitude as $k_\rho C$.}
Such a procedure should yield a completely stable sample
    temperature throughout the course of the ODNP experiment, and
    negate any concerns over changing sample temperature.
The current study also points out the clear benefit of
    the development of an error
    analysis that properly accounts for,
    among other detailed effects,
    the correlated error present in the $E(p)$ data,
    as well as the development of
    software and hardware to more fully automate ODNP
    measurements.
In particular, this data indicates that
    the primary bottleneck for quantitative accuracy
    in ODNP is the NMR relaxation (\ie $T_1$) measurements;
    a detailed statistics would help to highlight this fact,
    and more importantly, help to identify experimental strategies
    where the impact of such errors on the ultimate value
    of \Dlocal were minimized.
Finally, while previous studies could distinguish
    between surface, intermediate, and buried chemical sites
    on different systems,
    or compare the modulation in hydration
    dynamics within the same system,
    the new methodology presented here
    opens up the possibility of classifying
    different types of surface or buried
    sites based on more subtle differences
    in their local hydration dynamics reproducibly,
    \ie even across different studies and instruments.
\bibliography{library}
\end{document}